\documentclass[letterpaper,twocolumn,10pt]{article}

\usepackage{zhanggroup}
\usepackage{tikz}
\usepackage{amsfonts}
\usepackage{xspace} 
\usepackage{amsmath}
\usepackage{mathrsfs}
\usepackage{amssymb}
\usepackage{amsmath}
\usepackage{amsthm}
\usepackage{subcaption}
\usepackage{booktabs}
\hypersetup{
  colorlinks,
  linkcolor={blue!70!green},
  citecolor={green!70!blue},
  urlcolor={orange!70!red}
}

\newcommand{\mypara}[1]{\medskip\noindent{\bf{#1}.}\xspace}
\newcommand{\Dataset}{\mathcal{D}}
\newcommand{\TargetDataset}{\mathcal{D}_{\textit{Target}}}
\newcommand{\ShadowDataset}{\mathcal{D}_{\textit{Shadow}}}
\newcommand{\Shadow}{_{\textit{Shadow}}}
\newcommand{\Target}{_{\textit{Target}}}
\newcommand{\Train}{^{\textit{Train}}}
\newcommand{\Test}{^{\textit{Test}}}
\newcommand{\Graph}{\mathcal{G}}
\newcommand{\Subgraph}{g}
\newcommand{\NodeSet}{\mathcal{V}}
\newcommand{\EdgeSet}{\mathcal{E}}
\newcommand{\NodeFeatureSet}{\mathcal{X}}
\newcommand{\NodeLabelSet}{\mathcal{Y}}
\newcommand{\TargetNode}{v}
\newcommand{\Edge}{e}
\newcommand{\Radius}{l}
\newcommand{\GNNLayer}{t}

\newcommand{\NodeFeature}{x}
\newcommand{\NodeLabel}{y}
\newcommand{\Posterior}{p}
\newcommand{\AggregateFunction}{\mathsf{AGGREGATE}}
\newcommand{\UpdateFunction}{\mathsf{UPDATE}}
\newcommand{\RepresentationVector}{h}
\newcommand{\HiddenStateVector}{z}
\newcommand{\TargetModel}{\mathcal{M}_{\textit{T}}}
\newcommand{\ShadowModel}{\mathcal{M}_{\textit{S}}}
\newcommand{\AttackModel}{\mathcal{A}}
\newcommand{\AttackKnowledge}{\mathcal{K}}
\begin{document}
% ----------------------------------------------------

\date{}

\title{\Large \bf Node-Level Membership Inference Attacks Against Graph Neural Networks}

\author{
Xinlei He\textsuperscript{1}\ \ \
Rui Wen\textsuperscript{1}\ \ \
Yixin Wu\textsuperscript{2}\ \ \
Michael Backes\textsuperscript{1}\ \ \
Yun Shen\textsuperscript{3}\ \ \
Yang Zhang\textsuperscript{1}
\\
\\
\textsuperscript{1}\textit{CISPA Helmholtz Center for Information Security}\ \ \ 
\\
\textsuperscript{2}\textit{Sichuan University}\ \ \
\textsuperscript{3}\textit{NortonLifeLock Research Group}\ \ \

}

\maketitle

% ----------------------------------------------------
\begin{abstract}
% ----------------------------------------------------

Many real-world data comes in the form of graphs, such as social networks and protein structure. 
To fully utilize the information contained in graph data, a new family of machine learning (ML) models, namely graph neural networks (GNNs), has been introduced.
Previous studies have shown that machine learning models are vulnerable to privacy attacks.
However, most of the current efforts concentrate on ML models trained on data from the Euclidean space, like images and texts.
On the other hand, privacy risks stemming from GNNs remain largely unstudied.

In this paper, we fill the gap by performing the first comprehensive analysis of node-level membership inference attacks against GNNs.
We systematically define the threat models and propose three node-level membership inference attacks based on an adversary's background knowledge.
Our evaluation on three GNN structures and four benchmark datasets shows that GNNs are vulnerable to node-level membership inference even when the adversary has minimal background knowledge.
Besides, we show that graph density and feature similarity have a major impact on the attack's success.
We further investigate two defense mechanisms and the empirical results indicate that these defenses can reduce the attack performance but with moderate utility loss.

% ----------------------------------------------------
\end{abstract}
% ----------------------------------------------------

% ----------------------------------------------------
\section{Introduction}
% ----------------------------------------------------

Many real-world data can be organized in the form of graphs, such as social relations and protein structure.
Effective graph analysis provides users a deeper understanding of what is behind the data and can help to analyze many natural phenomena and build powerful commercial applications.
However, it is not trivial to utilize the classical machine learning models to analyze relational data with a more complex structure.
These models, such as convolutional neural networks (CNNs) and recurrent neural networks (RNNs), are designed to extract fine-grained representation for each data sample from its own feature.

To fully utilize the rich information of graph data, a new family of machine learning (ML) models, namely graph neural networks (GNNs)~\cite{KW17,HYL17,VCCRLB18,XHLJ19}, has been introduced to address graph-related tasks in an end-to-end manner.
GNN models utilize both the feature of each sample (referred to as a node in the GNN context) and the features of the sample's neighborhood (from the graph) to represent the sample.
In this way, a GNN learns to embed the structural connections among different nodes in its training dataset.

Recent research has shown that ML models are vulnerable to privacy attacks~\cite{SSSS17,SZHBFB19,FJR15,LF20,CLEKS19,MSCS19,SS19,SS20,CYZF20,CTWJHLRBSEOR20}.
Most of the current efforts in this direction concentrate on ML models trained on sensitive data from the Euclidean space, such as images and texts.
Meanwhile, graph data, which is used to train GNNs, also contains sensitive information, such as social relations~\cite{BDK07,CBCSHK10,JWZG17} and mobility traces~\cite{CML11,BHPZ17}.
However, the potential privacy risks stemming from GNNs have been largely understudied.

% ----------------------------------------------------
\subsection{Our Contributions} 
% ----------------------------------------------------

In this paper, we investigate whether a GNN model is vulnerable to membership inference attacks~\cite{SSSS17,SZHBFB19,LF20,SS19,HMDC19,CYZF20}, the major means to assess ML models' privacy risks.
Specifically, an adversary aims to infer whether a target node is used in the training dataset of a target GNN.
We concentrate on black-box membership inference, the most difficult setting for the adversary~\cite{SSSS17}.

There exists some preliminary work on privacy risks of GNNs~\cite{DBS20,ONK21,HJBGZ21}.
For instance, He et al.~\cite{HJBGZ21} propose a link stealing attack to infer the graph structure of a trained GNN model.
Also, Duddu et al.~\cite{DBS20} and Olatunji et al.~\cite{ONK21} have performed some preliminary studies on node-level membership inference attacks against GNNs.
However, the former lacks a clear attack methodology, while the latter conducts attacks in a restricted scenario, i.e., using a target node's 2-hop subgraph to query the target model to obtain the input to their attack model (see \autoref{subsection:AttackMethodology}), which falls short of providing a complete picture of GNN's membership inference risks.

As mentioned before, GNNs are designed for graph data that is not in the Euclidean space, which leads to some unique research questions for membership inference attacks in this setting.
First, an adversary needs background knowledge, such as the target GNN's architecture and a shadow dataset, to train their attack model.
State-of-the-art GNN models are normally shallow with less diverse choices of model architectures compared to CNNs and RNNs due to the fact that real-world graphs normally exhibit small-world phenomenon~\cite{EK10}. 
Also, different graphs share many common properties, such as power-law degree distribution~\cite{LRU14}.
This motivates us to understand whether an adversary can have less constrained background knowledge compared to previous membership inference attacks against other types of ML models.
Second, an adversary can query a target node to a target GNN with either the node's feature alone or the node and its neighborhood's graph connections as well as their features.
This means one node can receive two different prediction outputs (posteriors) from the target GNN.
We are interested in which posteriors reveal more information of the target node's membership status and whether these two posteriors can be combined to achieve a more effective attack.
To answer these research questions, we make the following contributions in this paper.

We first systematically define the threat model of node-level membership inference attack against GNNs by categorizing an adversary's background knowledge along three dimensions, i.e., shadow dataset, shadow model, and node topology.
Specifically, we assume that an adversary may have a shadow dataset that comes from the same or different distribution of the target model's training dataset.
They can also establish a shadow model that has the same or different architecture from the target model.
Regarding node topology, we consider two situations: 1) the adversary only uses the target node's feature itself to query the target model (0-hop query) or 2) the adversary uses the target node and its 2-hop subgraph's information to query the target model (2-hop query). 

Following the different threat models based on node topology, we propose three membership inference attack models, namely 0-hop attack, 2-hop attack, and combined attack.
We perform an extensive evaluation on three popular GNN models including GraphSAGE~\cite{HYL17}, Graph Attention Network (GAT)~\cite{VCCRLB18}, and Graph Isomorphism Network (GIN)~\cite{XHLJ19} with four benchmark datasets, i.e., Cora~\cite{KW17}, Citeseer~\cite{KW17}, Cora-full~\cite{BG182}, and LastFM Asia~\cite{RS20}.
Experimental results show that our attacks achieve strong performance.
For instance, our 0-hop attack achieves 0.791 accuracy on the GraphSAGE model trained on Citeseer.
More interestingly, we discover that our 0-hop attack has better performance than the 2-hop attack.
This is due to the fact a target node's 2-hop neighborhood contains a mixture of member and non-member nodes which jeopardizes the attack model's accuracy.
Also, our combined attack achieves the strongest performance by taking advantage of both the 0-hop and 2-hop attacks.
Moreover, we show that when the adversary does not know the target model's training dataset distribution or architecture, our attack is still effective.

We perform an in-depth analysis of the success behind the attack.
Our experiments reveal that a target node with higher subgraph density is more prone to membership inference.
This is due to the fact that a dense subgraph drives the node to participate more in the aggregation process of the GNN training, which amplifies the node's influence in the target GNN model.
Besides, it is easier for the adversary to mount their attack if a node shares similar features with its neighbors.

We propose two defense mechanisms to mitigate the membership inference risks of GNNs.
Empirical evaluation shows that they are able to mitigate the attack performance to a certain extent while bringing moderate utility damage.
This motivates us to further investigate advanced defenses in the future.

% ----------------------------------------------------
\subsection{Organization}
% ----------------------------------------------------

The rest of the paper is organized as the following.
In \autoref{section:gnn}, we provide some background knowledge of graph neural networks.
\autoref{section:MIAAgainstGNN} presents the threat model and attack methodology.
In \autoref{section:Evaluation}, we discuss our empirical evaluation results.
\autoref{section:RelatedWork} summarizes the related work and \autoref{section:Conclusion} concludes the paper.

% ----------------------------------------------------
\section{Graph Neural Networks}
\label{section:gnn}
% ----------------------------------------------------

In this section, we first introduce the notations used in the paper.
Then, we introduce the three representative GNN architectures we focus on.
In the end, we discuss the prediction process of GNN.

% ----------------------------------------------------
\subsection{Notations}
% ----------------------------------------------------

We define a graph dataset as $\Dataset=(\Graph, \NodeFeatureSet, \NodeLabelSet)$.
Here, $\Graph=(\NodeSet, \EdgeSet)$ represents a graph with $\NodeSet$ denoting the graph's set of nodes and $\EdgeSet$ representing the set of edges connecting these nodes.
Each node is denoted by $\TargetNode \in \NodeSet$ and $\Edge_{uv} \in \EdgeSet$ represents an edge linking two nodes $u$ and $v$.
$\NodeFeatureSet=\{\NodeFeature_1,\NodeFeature_2,...,\NodeFeature_{\vert\NodeSet\vert}\}$ and $\NodeLabelSet= \{\NodeLabel_1, \NodeLabel_2,...,\NodeLabel_{\vert\NodeSet\vert}\}$ represent the features and labels for all the nodes in $\Graph$, respectively.
Node $\TargetNode$'s $\Radius$-hop neighborhood is denoted by $\mathcal{N}^{\Radius}({\TargetNode})$, which contains a set of nodes at a distance less than or equal to $\Radius$ from $\TargetNode$ in $\Graph$.
For convenience, we abbreviate the $1$-hop neighborhood of $\TargetNode$ as $\mathcal{N}({\TargetNode})$.
The $\Radius$-hop subgraph of node $\TargetNode$, denoted by $\Subgraph^{\Radius}(\TargetNode)$, contains $\TargetNode$ and its $\Radius$-hop neighborhood $\mathcal{N}^{\Radius}({\TargetNode})$, edges among these nodes, and features of these nodes.
We summarize the main notations in ~\autoref{table:notion}.

\begin{table}[!t]
\caption{List of notations.}
\label{table:notion}
\centering
\renewcommand{\arraystretch}{1.1}
\begin{tabular}{r|l}
\toprule
Notation  & Description \\
\midrule
$\Dataset =(\Graph,\NodeFeatureSet,\NodeLabelSet)$ &  Dataset\\
$\TargetNode$ & A node\\
$\mathcal{N}^{\Radius}({\TargetNode})$ & ${\Radius}$-hop neighborhood of $\TargetNode$\\
$\Subgraph^{\Radius}(\TargetNode)$ & ${\Radius}$-hop subgraph of $\TargetNode$\\
$\RepresentationVector_{v}^{(\GNNLayer)}$ &  Representation vector of $\TargetNode$ at layer $\GNNLayer$\\
$\TargetModel$ &  Target model\\
$\Dataset\Target$ &  Target dataset \\
$\Dataset\Target\Train$ ($\Dataset\Target\Test$) & Target training (testing) dataset \\
$\ShadowModel$ &  Shadow model\\
$\Dataset\Shadow$ &  Shadow dataset \\
$\Dataset\Shadow\Train$ ($\Dataset\Shadow\Test$) &  Shadow training (testing) dataset \\
\bottomrule
\end{tabular}
\end{table}

% ----------------------------------------------------
\subsection{GNN Architecture} 
\label{subsection:gnn_architecture}
% ----------------------------------------------------

In general, there are two settings for GNNs, i.e., \textit{transductive} setting and \textit{inductive} setting.
In the transductive setting, a GNN (e.g., vanilla GCN~\cite{KW17}) is trained and tested on the same fixed graph.
It means, in the testing phase, the GNN model can only provide predictions for nodes that are in its training dataset.
Therefore, membership inference attacks against transductive GNN models are trivial.
In this paper, we focus on the inductive setting of GNNs, whereby a GNN model can classify nodes that are not from its training dataset.

Basically, a GNN contains several graph convolution layers that iteratively update a node $\TargetNode$'s representation by aggregating the representation of nodes in $\TargetNode$'s neighborhood.
Formally, each graph convolution layer of a GNN model can be defined as follows:

\begin{equation}
\begin{array}{l}
\HiddenStateVector_{v}^{(\GNNLayer)} =\AggregateFunction(\{\RepresentationVector_{u}^{(\GNNLayer-1)}: u \in \mathcal{N}(v)\}) \\
\RepresentationVector_{v}^{(\GNNLayer)}=\UpdateFunction( \HiddenStateVector_{v}^{(\GNNLayer)})
\end{array}
\end{equation}
where $\mathcal{N}({v})$ is the neighborhood of $v$.
$\GNNLayer$ represents the $\GNNLayer$-th layer of the GNN.
$\HiddenStateVector_{v}^{(\GNNLayer)}$ and $\RepresentationVector^{(\GNNLayer)}_{v}$ denote the hidden state and the representation vector of node $v$ at layer $\GNNLayer$.
In the first step, we initialize $\TargetNode$'s representation $h_{v}^{(0)}$ as its feature $\NodeFeature_{\TargetNode}$.

$\AggregateFunction(\cdot)$ and $\UpdateFunction(\cdot)$ are the aggregation and update functions, respectively.
For a given node $v$, the aggregation function is used to generate the current hidden state $\HiddenStateVector_{v}^{(\GNNLayer)}$ using a combination of its previous representation and the aggregated representation from its neighborhood $\mathcal{N}(v)$.
The update function then conducts non-linear transformation on the current hidden state $\HiddenStateVector_{v}^{(\GNNLayer)}$ and produces the representation vector $\RepresentationVector^{(\GNNLayer)}_{v}$.
A multilayer perceptron (MLP) is usually used as the update function's structure.
Meanwhile, different GNN models may leverage different aggregation functions.
In this paper, we focus on three representative GNN architectures, i.e., GraphSAGE~\cite{HYL17}, Graph Attention Network (GAT)~\cite{VCCRLB18}, and Graph Isomorphism Network (GIN)~\cite{XHLJ19}.

\mypara{GraphSAGE} 
Hamilton et al.~\cite{HYL17} propose GraphSAGE, which first generalizes the original graph convolutional network~\cite{KW17} to the inductive setting with different aggregation functions.
In this paper, we follow the widely used mean aggregation function of GraphSAGE, which can be defined as follows:

\begin{equation}
\HiddenStateVector_{v}^{(\GNNLayer)} = \operatorname{\mathsf{CONCAT}}(\mathrm{h}_{v}^{(\GNNLayer-1)}, \frac{1}{|\mathcal{N}(v)|}\sum_{u \in \mathcal{N}(v)}\mathrm{h}_{u}^{(\GNNLayer-1)}) 
\end{equation}
where $\operatorname{\mathsf{CONCAT}}$ is the concatenation operation.

\mypara{GAT} 
Inspired by the attention mechanism in deep learning~\cite{VSPUJGKP17}, Velickovic et al.~\cite{VCCRLB18} propose GAT that leverages multi-head attention to learn different attention weights and pays more attention to important neighborhoods. 
GAT's aggregation function can be formulated as:

\begin{equation} 
\HiddenStateVector_{v}^{(\GNNLayer)} = \operatorname{\mathsf{CONCAT}}_{k=1}^{K} \sigma(\sum_{u \in \mathcal{N}(v)}\alpha_{uv}^{k} \mathbf{W}^{k} \mathrm{h}_{u}^{(\GNNLayer-1)}) 
\end{equation}

where $K$ is the total number of projection heads in the attention mechanism. $\mathbf{W}^{k}$ and $\alpha_{uv}^{k}$ are the weight matrix and the attention coefficient in the $k$-th projection head, respectively.
$\sigma(\cdot)$ is the activation function.

\mypara{GIN} 
Xu et al.~\cite{XHLJ19} develop GIN whose representation power is well-matched with the Weisfeiler-Lehman test for graph isomorphism.
The aggregation function of GIN can be represented as:
\begin{equation}
\HiddenStateVector_{v}^{(\GNNLayer)}=  (1+\epsilon^{(\GNNLayer)}) \cdot h_{v}^{(\GNNLayer-1)}+\sum_{u \in \mathcal{N}(v)} \mathrm{h}_{u}^{(\GNNLayer-1)}
\end{equation}
where $\epsilon$ is a learnable parameter to adjust the weight of node $\TargetNode$.

% ----------------------------------------------------
\subsection{GNN Prediction}
% ----------------------------------------------------

In this paper, we focus on node classification tasks.
In the training phase, an inductive GNN learns the parameters of aggregation and update functions in different layers over a training dataset.
Then, to get a precise prediction of an unseen node $\TargetNode$ in a $\GNNLayer$-layer GNN, we can feed $\TargetNode$'s $\GNNLayer$-hop subgraph, i.e., $\Subgraph^{\GNNLayer}(\TargetNode)$, to the GNN and obtain the prediction posteriors $\Posterior_{\TargetNode}$.
Note that the $\GNNLayer$-hop subgraph of $\TargetNode$ is not a necessary condition to acquire the posteriors $\Posterior_{\TargetNode}$.
We can obtain posteriors $\Posterior_{\TargetNode}$ by only querying the target node $\TargetNode$'s feature to the GNN model.
Our evaluation shows that even in this case, the GNN model can achieve better performance than MLP, i.e., a model that does not consider graph structural information (see \autoref{section:Evaluation}).

% ----------------------------------------------------
\section{Node-Level Membership Inference Against GNNs}
\label{section:MIAAgainstGNN}
% ----------------------------------------------------

In this section, we first define node-level membership inference attacks against GNNs.
Then, we discuss the threat model and present the attack methodology.

% ----------------------------------------------------
\subsection{Problem Definition}
\label{subsection:ProblemDefinition}
% ----------------------------------------------------

The goal of an adversary is to determine whether a given node is used to train a target GNN model or not.
More formally, given a target node $\TargetNode$, a target GNN model $\TargetModel$, and the adversary's background knowledge $\AttackKnowledge$, node-level membership inference attack $\AttackModel$ is defined as the following.

\begin{equation}
\AttackModel: \TargetNode, \TargetModel, \AttackKnowledge \mapsto \{\textit{member}, \textit{non-member}\}  
\end{equation}
As discussed in the previous work~\cite{SSSS17}, successful membership inference attacks can cause severe privacy risks.
In the setting of GNNs, membership threat is related to graph data, such as inferring a user being a member of a sensitive social network.

\begin{figure*}[!t]
\centering
\includegraphics[width=2\columnwidth]{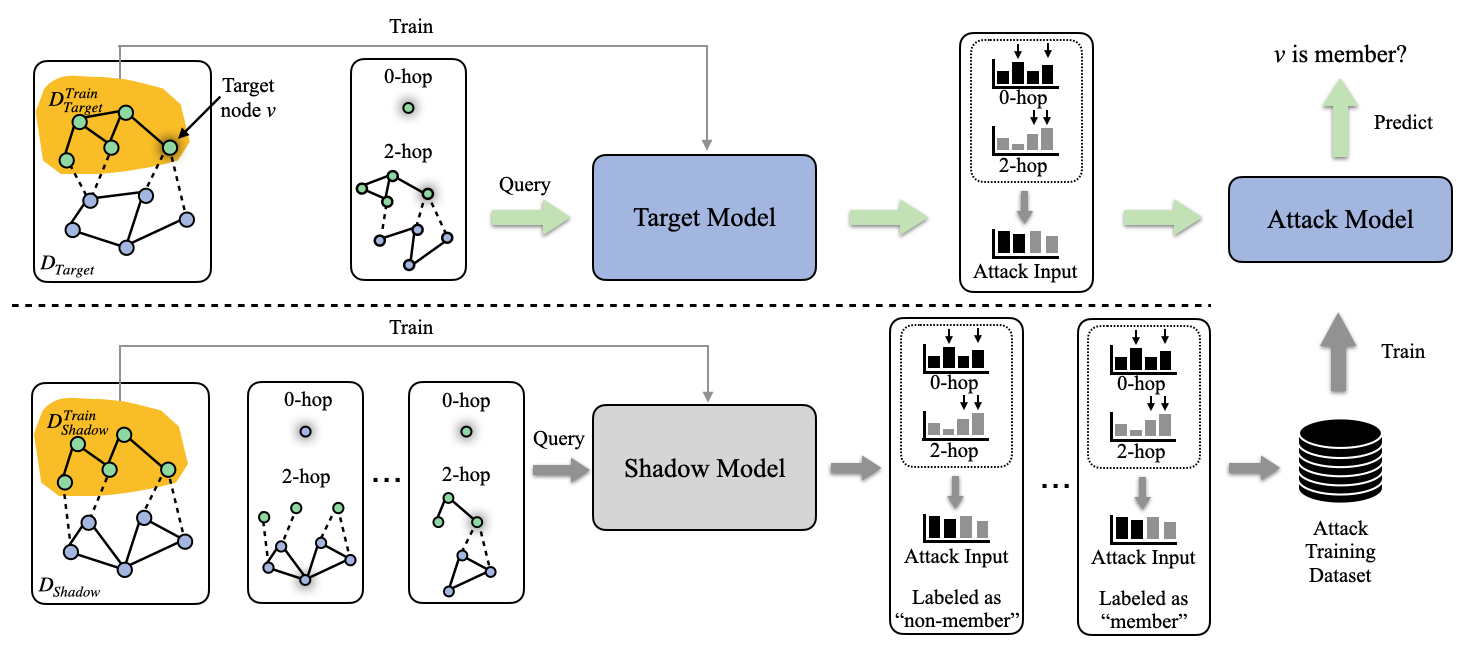}
\caption{A schematic overview of node-level membership inference attack against GNNs.}
\label{figure:attack_pipeline}
\end{figure*} 

% ----------------------------------------------------
\subsection{Threat Model}
\label{subsection:ThreatModel}
% ----------------------------------------------------

Our target model $\TargetModel$ is an inductive GNN model. 
First, we assume that the adversary only has black-box access to the target model, i,e, they can only query the target model and obtain the posteriors.
As mentioned by previous work~\cite{SSSS17,SZHBFB19,HJBGZ21}, black-box setting is the most challenging scenario for the adversary.
We then categorize the adversary's background knowledge $\AttackKnowledge$ along three dimensions, i.e., shadow dataset, shadow model, and node topology.

\mypara{Shadow Dataset} 
We assume that the adversary has a shadow dataset $\ShadowDataset$ which contains its own graph structure as well as node features and labels.
Following the previous work~\cite{SSSS17}, the shadow dataset $\ShadowDataset$ can come from the same distribution of the target model's training dataset.
However, our empirical evaluation shows that this assumption can be relaxed (see \autoref{subsection:EvaluationRelaxAssumptions}).
Note that in both cases, the shadow dataset has no node and edge intersection with the target dataset.

\mypara{Shadow Model}
With the shadow dataset, the adversary can train a shadow GNN model $\ShadowModel$ to mimic the behaviors of the target model $\TargetModel$.
We can assume that the shadow model shares the same architecture as the target model~\cite{SSSS17,SZHBFB19}.
In this case, the adversary needs to first perform a hyperparameter stealing attack to obtain the target model's architecture~\cite{WG18}.
Also, our experimental results show that an adversary can use a different GNN architecture from the target model to establish their shadow model (see \autoref{subsection:EvaluationRelaxAssumptions}).

\mypara{Node Topology}
To get the posteriors for $\TargetNode$ from $\TargetModel$, we consider two cases.
In the first case, we assume that the adversary only has $\TargetNode$'s feature $\NodeFeature_\TargetNode$.
As the input to a GNN needs to be in the form of a graph, we add a self-loop for $\TargetNode$~\cite{KW17} and query the target model.
We refer to this case as a node's \textit{0-hop query}.
In the second case, we assume that the adversary knows the target node $\TargetNode$'s 2-hop subgraph $\Subgraph^{2}(\TargetNode)$, which can be directly fed to the target model.
We name this scenario as a node's \textit{2-hop query}.
Note that $\Subgraph^{2}(\TargetNode)$ does not need to be the complete 2-hop subgraph of $\TargetNode$ as the adversary may only have a partial view of the dataset.
Besides, nodes in $\Subgraph^{2}(\TargetNode)$ can be a mixture of members and non-members for the target GNN.
This is more realistic as the adversary does not know any other nodes' membership status.
The goal is to infer the membership status of $\TargetNode$.
In this paper, we only consider the 0-hop and 2-hop queries since they are the two extreme querying cases where 0-hop query utilizes no information from the graph structure, while 2-hop query considers the complete graph structure (2-hop subgraph).\footnote{Most of the state-of-the-art GNNs follow two-layer structure due to the fact that real-world graphs normally exhibit small-world phenomenon~\cite{EK10}, and in this case, 2-hop query is the upper bound for the query depth.
Moreover, previous empirical results~\cite{HYL17} show that deeper GNN architecture does not further improve the classification performance.}
Indeed, 1-hop subgraph is also a possible node topology to the adversary.
We leave the investigation as our future work.

% ----------------------------------------------------
\subsection{Attack Methodology}
\label{subsection:AttackMethodology}
% ----------------------------------------------------

Following the standard process of membership inference attacks against ML models~\cite{SSSS17}, our attack can be divided into three stages, i.e., shadow model training, attack model training, and membership inference.
\autoref{figure:attack_pipeline} provides a schematic overview of the attack process.

\mypara{Shadow Model Training}
Given a shadow dataset $\ShadowDataset$, the adversary first splits its node set $\NodeSet\Shadow$ into two disjoint sets, including $\NodeSet\Shadow\Train$ and $\NodeSet\Shadow\Test$.
Then, the adversary derives their shadow training ($\ShadowDataset\Train$) and testing ($\ShadowDataset\Test$) datasets by involving all the features, labels, and edges within $\NodeSet\Shadow\Train$ and $\NodeSet\Shadow\Test$, respectively.
After that, $\ShadowDataset\Train$ is used to train a shadow GNN model $\ShadowModel$.

\mypara{Attack Model Training}
The attack model is a binary machine learning classifier and its input is derived from a node's posteriors provided by a GNN.
To obtain the training dataset for the attack model, the adversary needs to query $\ShadowModel$ with all the nodes in $\NodeSet\Shadow$ (both $\NodeSet\Shadow\Train$ and $\NodeSet\Shadow\Test$) and gets the corresponding prediction posteriors.
As mentioned before, depending on their knowledge of node topology, the adversary can perform 0-hop query or 2-hop query.
For a node $\TargetNode$, we refer to its posteriors obtained by 0-hop query (2-hop query) as \textit{0-hop posteriors} (\textit{2-hop posteriors}).
In this paper, we consider three types of attack model input summarized from posteriors which leads to three attack models, namely 0-hop attack~$\AttackModel_{0}$, 2-hop attack~$\AttackModel_{2}$, and combined attack~$\AttackModel_{c}$.
\begin{itemize}
\item \textbf{0-hop Attack.}
The 0-hop attack model is essentially an MLP, which takes $\TargetNode$'s largest two\footnote{Classification tasks considered in this paper have at least two classes.} values (ranked) in its 0-hop posteriors as the input.
\item \textbf{2-hop Attack.}
The 2-hop attack model is also an MLP, which takes $\TargetNode$'s largest two values (ranked) in its 2-hop posteriors as the input.
\item \textbf{Combined Attack}
The combined attack model considers both the inputs for the 0-hop and the 2-hop attack by first feeding them separately to different linear layers.
Then, the attack model concatenates the two embeddings and feeds them to an MLP. 
\end{itemize}
Note that if the adversary can perform 2-hop attack of a given node, they can also perform 0-hop attack.
Therefore, the combined attack requires the same background knowledge as the 2-hop attack.
In all cases, if $\TargetNode\in\NodeSet\Shadow\Train$, we label it as a member, otherwise as a non-member.
In the end, the adversary constructs an attack training dataset, which they use to train their attack model.

\mypara{Membership Inference}
To determine whether a target node is used to train the target model $\TargetModel$, the adversary first conducts 0-hop query or 2-hop query to the target model depending on their background knowledge. 
Then, the adversary queries the attack model with the 0-hop posteriors, 2-hop posteriors, or both to get the node's membership prediction.

% ----------------------------------------------------
\section{Evaluation}
\label{section:Evaluation}
% ----------------------------------------------------

In this section, we perform a comprehensive measurement of the node-level membership privacy risks stemming from GNN models.
We first introduce the experimental setup, then present the evaluation results for the attacks in different settings.
In the end, we evaluate the performance of possible defense mechanisms.

% ----------------------------------------------------
\subsection{Experimental Setup}
% ----------------------------------------------------

\mypara{Dataset}
We conduct experiments on four public datasets, including Cora~\cite{KW17}, Citeseer~\cite{KW17}, Cora-full~\cite{BG182}, and LastFM Asia~\cite{RS20} (abbreviated as Lastfm).
Cora and Citeseer are citation graphs whose nodes represent papers and edges reflect citation relationships among papers. 
Cora-full is an extended Cora dataset.
Lastfm is a social network dataset with its nodes being users and edges representing users' mutual following relationships. 
All datasets contain node features and labels.
Dataset statistics are summarized in \autoref{table:dataset_statistics}.

\begin{table}[!t]
\caption{Dataset statistics.}
\label{table:dataset_statistics}
\centering
\renewcommand{\arraystretch}{1.1}
\begin{tabular}{c|c c c c}
\toprule
Dataset  & \#. Node & \#. Edge & \#. Feature & \#. Class \\
\midrule
Cora & 2,708 & 5,429 & 1,433 & 7 \\
Citeseer & 3,327 & 4,732 & 3,703 & 6\\
Cora-full & 19,793 & 65,311 & 8,710 & 70\\
Lastfm & 7,624 & 27,806 & 7,842 & 18\\
\bottomrule
\end{tabular}
\end{table}

\mypara{Dataset Configuration}
The dataset configuration process is depicted in \autoref{figure:attack_pipeline}.
For each dataset, we first randomly split its nodes by half.
The first half (including the nodes, the edges among the nodes, and the nodes' features and labels) is used to construct the target dataset, i.e., $\TargetDataset$.
The other half is treated as the shadow dataset, i.e., $\ShadowDataset$.
Note that the target dataset and shadow dataset are disjoint as mentioned in \autoref{section:MIAAgainstGNN}.
For the target dataset $\TargetDataset$, we further randomly split it by half creating the target training dataset $\TargetDataset\Train$ and the target testing dataset $\TargetDataset\Test$. 
The target training dataset is used to train the target model, and the target testing dataset is used to test the target model's performance with respect to its original classification task.
Both $\TargetDataset\Train$ and $\TargetDataset\Test$ are used to test membership inference. 
Nodes in $\TargetDataset\Train$ are considered as members and nodes in $\TargetDataset\Test$ as non-members.
As mentioned in \autoref{subsection:ThreatModel}, for the 2-hop query scenario, each node's 2-hop subgraph can contain a mixture of member and non-member nodes.

We apply the same processing procedure on the shadow dataset to generate the shadow training dataset $\ShadowDataset\Train$ and the shadow testing dataset $\ShadowDataset\Test$.
$\ShadowDataset\Train$ is used to train the shadow model.
Both $\ShadowDataset\Train$ and $\ShadowDataset\Test$ are used to derive the training dataset for the attack model.

\mypara{Metric}
We use accuracy as our evaluation metric for both target model's performance and attack model's performance as it is widely used in node classification tasks~\cite{KW17,VCCRLB18,XHLJ19} as well as membership inference attacks~\cite{SSSS17,SZHBFB19}. 

\mypara{Target Models}
We leverage three GNN architectures, i.e., GraphSAGE, GAT, and GIN, to construct our target models and shadow models.
For each target model, we set the number of layers to 2 and the number of neurons to 32 in the hidden layer. 
Additionally, GAT models require the specification of the number of heads in the multi-head attention mechanism. 
We set the number of heads for the first layer and the second layer to 2 and 1, respectively. 
We also use dropout in all hidden layers to reduce overfitting, and the dropout rate is 0.5. 
We adopt cross-entropy as the loss function and Adam as the optimizer.
The learning rate is set to 0.003. 
The target and shadow models are both trained for 200 epochs.

\mypara{Baseline Model}
We leverage a 2-layer MLP as the baseline model to perform the same task as the target model's original task. 
Each hidden layer has 32 neurons with ReLU as its activation function. 
Loss function, optimizer, epochs, and learning rate are identical to those of the target GNN models.

\mypara{Attack Models}
For both 0-hop and 2-hop attacks, a 2-layer MLP is utilized as the attack model and the number of neurons in the hidden layer is set to 128.
Regarding the combined attack, the two inputs are first fed into two separated linear layers (with 64 neurons) simultaneously.
We then concatenate the two 64-dimensional embeddings and feed them to another linear layer for membership inference.
ReLU is adopted as the activation function for all the attack models.
Also, the loss function and optimizer are the same as the target model.
We set the learning rate to 0.001 and the training epochs to 500.

\mypara{Implementation}
Our code is implemented with PyTorch\footnote{\url{https://pytorch.org/}} and DGL.\footnote{\url{https://www.dgl.ai}}
The experiments are performed on an NVIDIA DGX-A100 server with Ubuntu 18.04 system.

\begin{figure}[!t]
\centering
\includegraphics[width=\columnwidth]{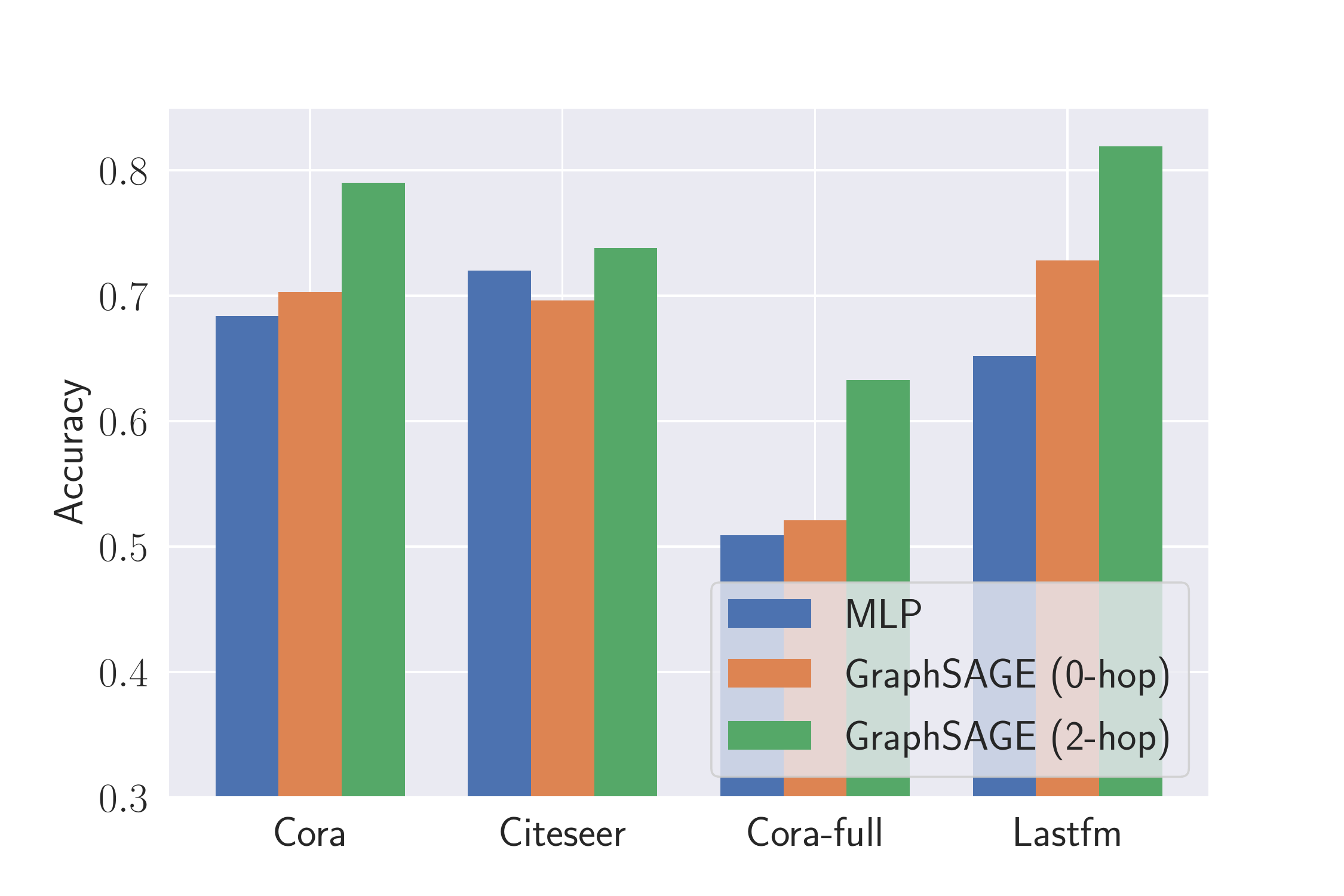}
\caption{The performance of original classification tasks when the target model's architecture is MLP or GraphSAGE (0-hop and 2-hop query).
The x-axis represents different datasets. 
The y-axis represents original classification tasks' accuracy.}
\label{figure:target_model_performance}
\end{figure} 

\begin{figure*}[!t]
\centering
\begin{subfigure}{0.50\columnwidth}
\includegraphics[width=\columnwidth]{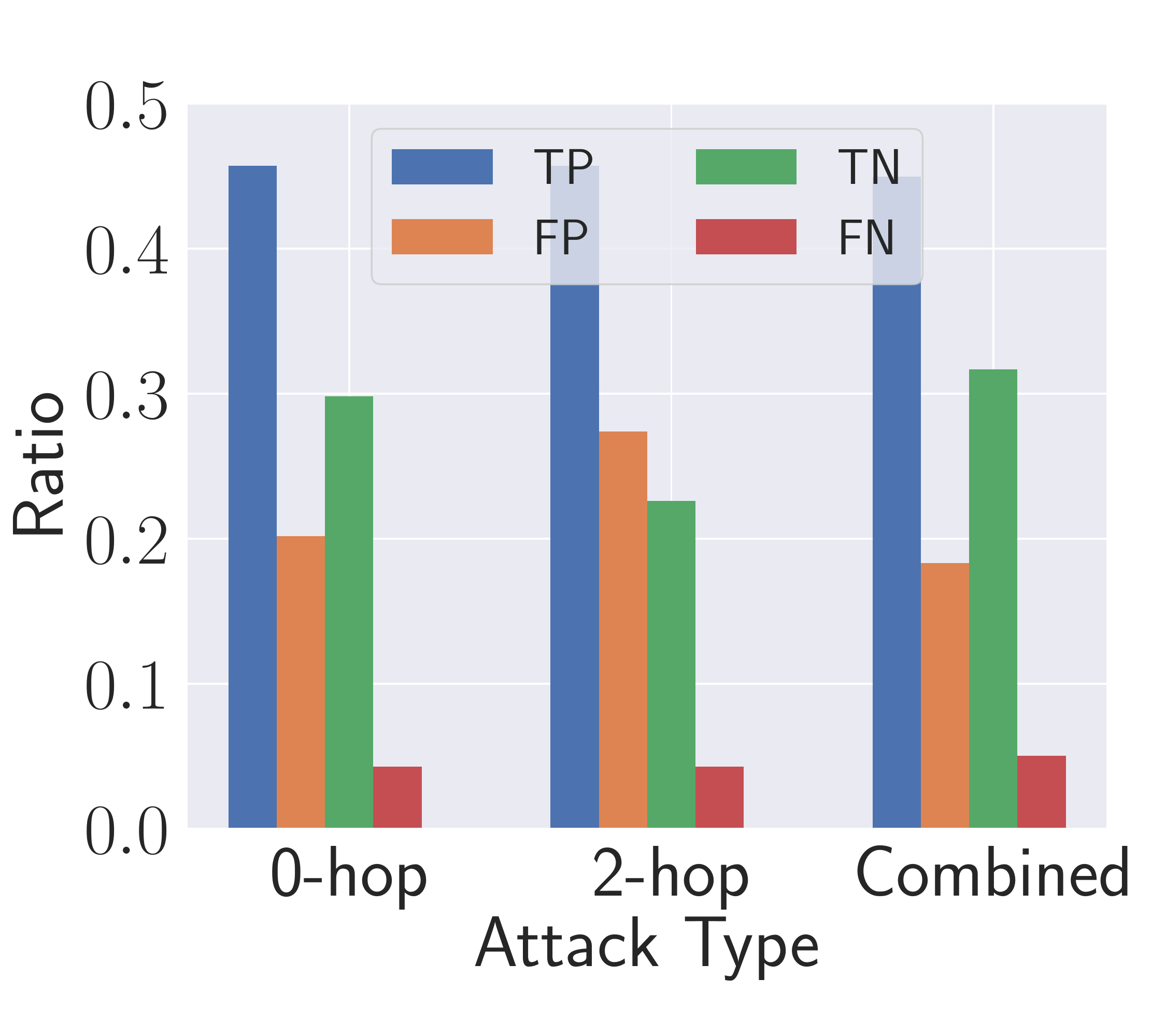}
\caption{Cora}
\label{figure:tpfp_analy_Cora_log}
\end{subfigure}
\begin{subfigure}{0.50\columnwidth}
\includegraphics[width=\columnwidth]{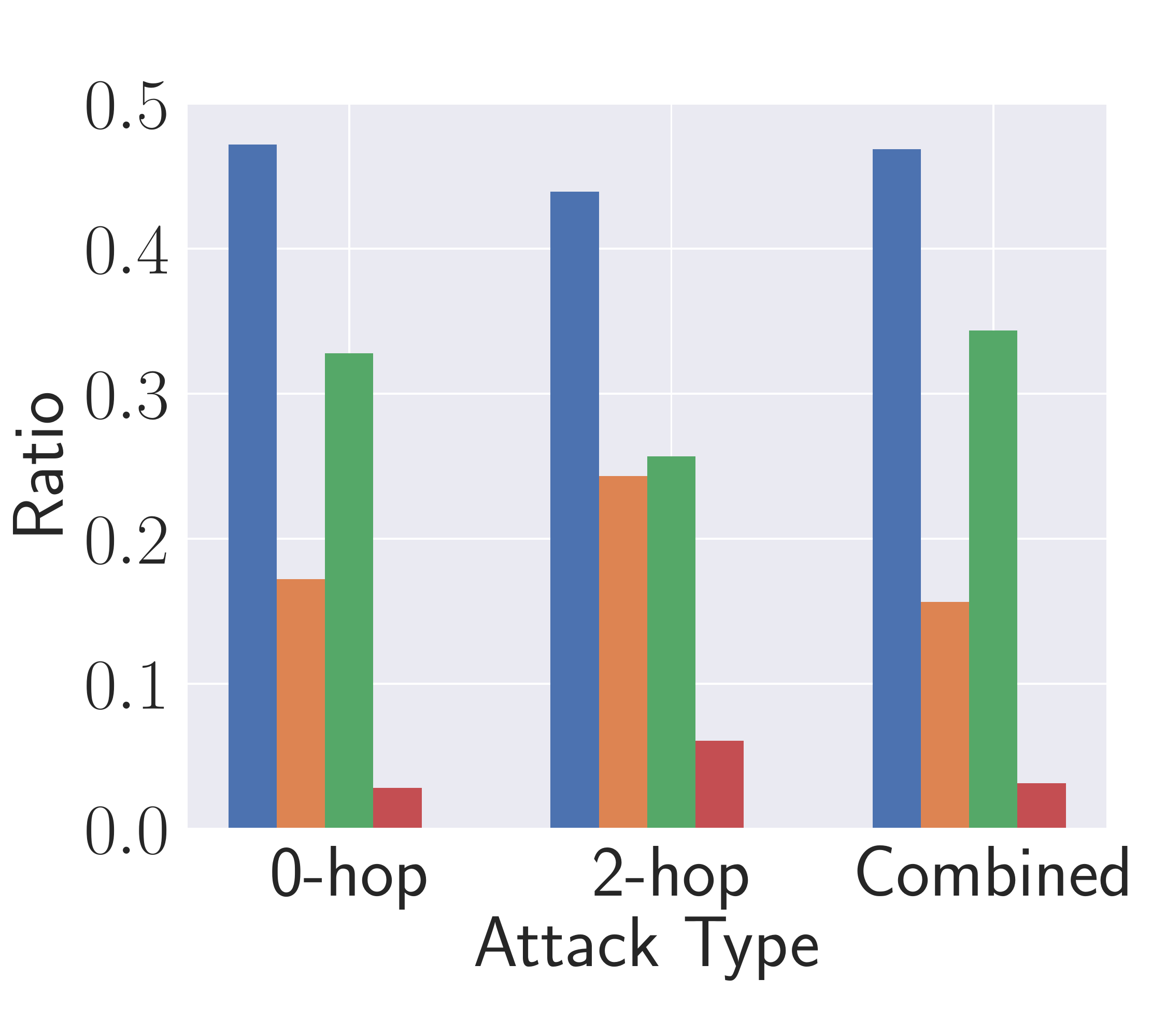}
\caption{Citeseer}
\label{figure:tpfp_analy_Citeseer_log}
\end{subfigure}
\begin{subfigure}{0.50\columnwidth}
\includegraphics[width=\columnwidth]{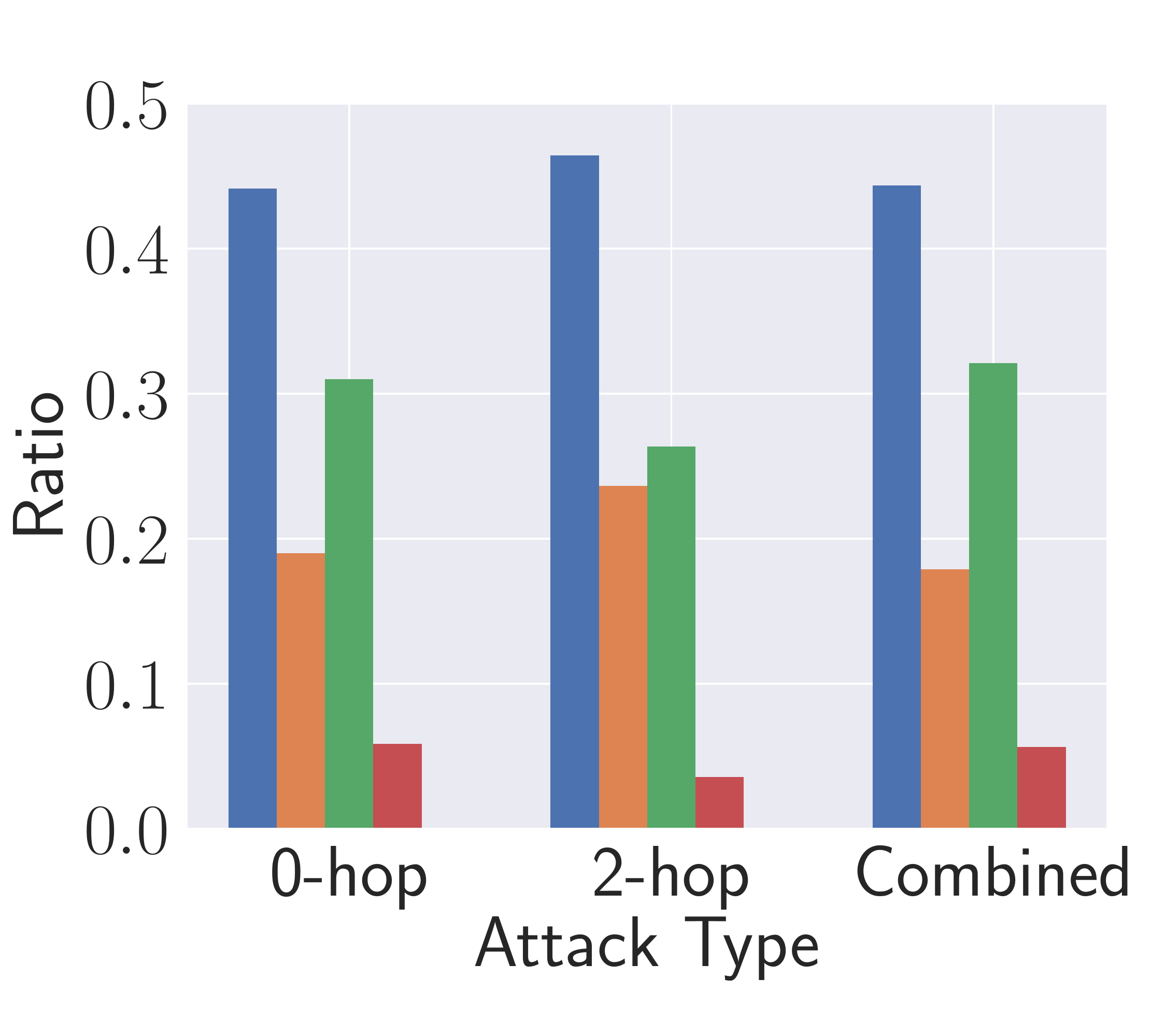}
\caption{Cora-full}
\label{figure:tpfp_analy_Cora-full_log}
\end{subfigure}
\begin{subfigure}{0.50\columnwidth}
\includegraphics[width=\columnwidth]{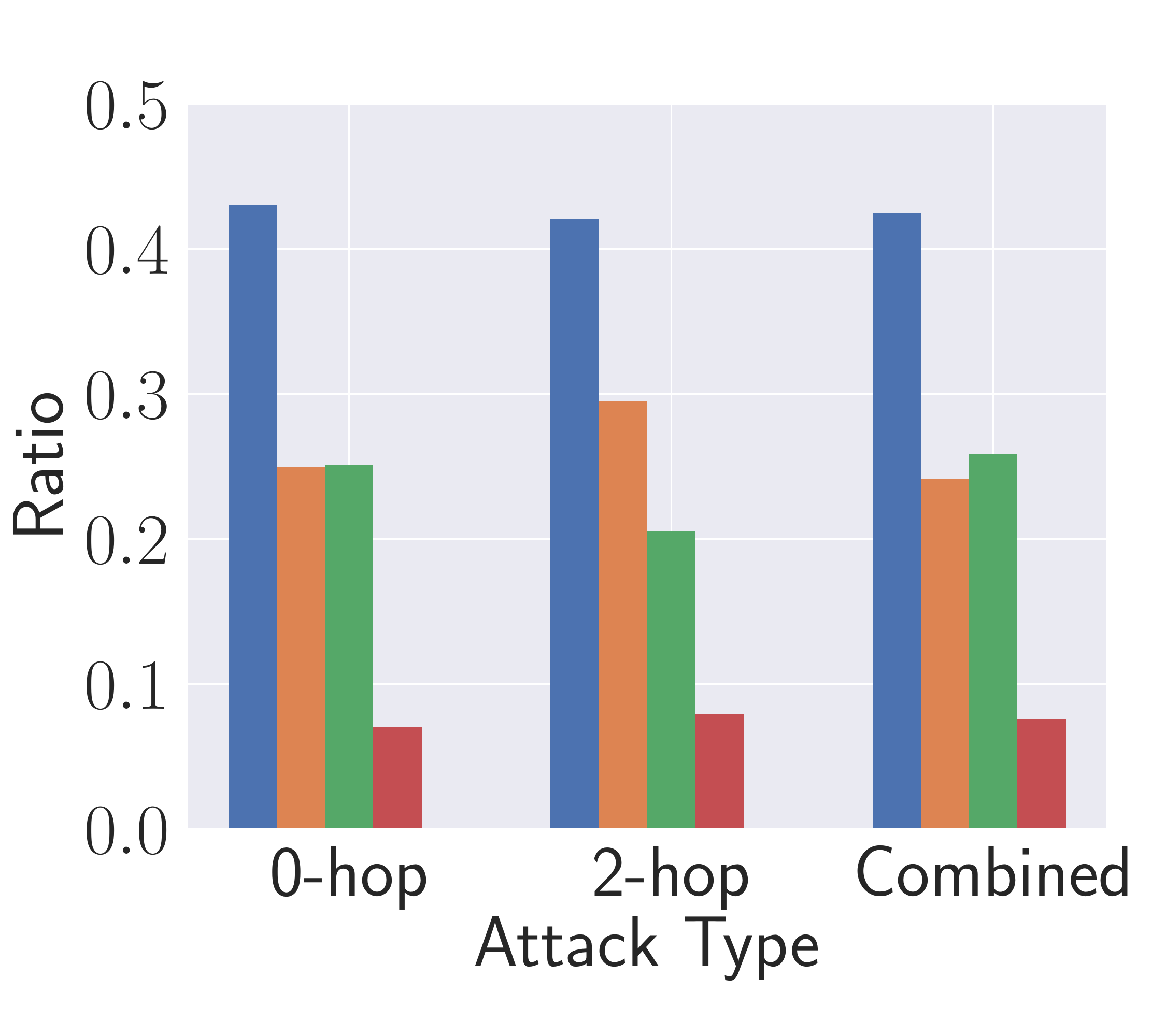}
\caption{Lastfm}
\label{figure:tpfp_analy_Lastfm_log}
\end{subfigure}
\caption{The ratio of true positive (TP), false positive (FP), true negative (TN), and false negative (FP) of different attacks for GraphSAGE on four different datasets.
The x-axis represents different attack types.
The y-axis represents the ratio.}
\label{figure:tpfp_analysis}
\end{figure*}

% ----------------------------------------------------
\subsection{Target Model Performance}
\label{sec:target_model_performance}
% ----------------------------------------------------

We first show the performance of the target models with respect to their original classification tasks in \autoref{figure:target_model_performance}.
To get the posteriors of a given node, we consider two query scenarios for each target model, i.e., 0-hop query and 2-hop query.
For comparison, we only consider a node's feature as the input to each baseline model (i.e., a 2-layer MLP) and perform the same classification task as the target model.

Due to space limitations, we only show the results for GraphSAGE.
Other GNN models exhibit similar trends.
First of all, compared to MLP, we observe that GNN has higher performance in the original task when using 2-hop queries.
For instance, on the Cora dataset, the baseline MLP achieves 0.684 accuracy while the GraphSAGE (2-hop) achieves 0.790 accuracy.
This demonstrates the efficacy of GNN models that consider nodes' features as well as their neighborhood information jointly for classification.
Second and more interestingly, 0-hop query on GraphSAGE also achieves better performance than MLP except for Citeseer.
This indicates that the graph information used during the training phase can be generalized to boost the performance of a GNN model even when it is queried with only a node's feature (0-hop query).

\begin{figure*}[!t]
\centering
\begin{subfigure}{0.66\columnwidth}
\includegraphics[width=\columnwidth]{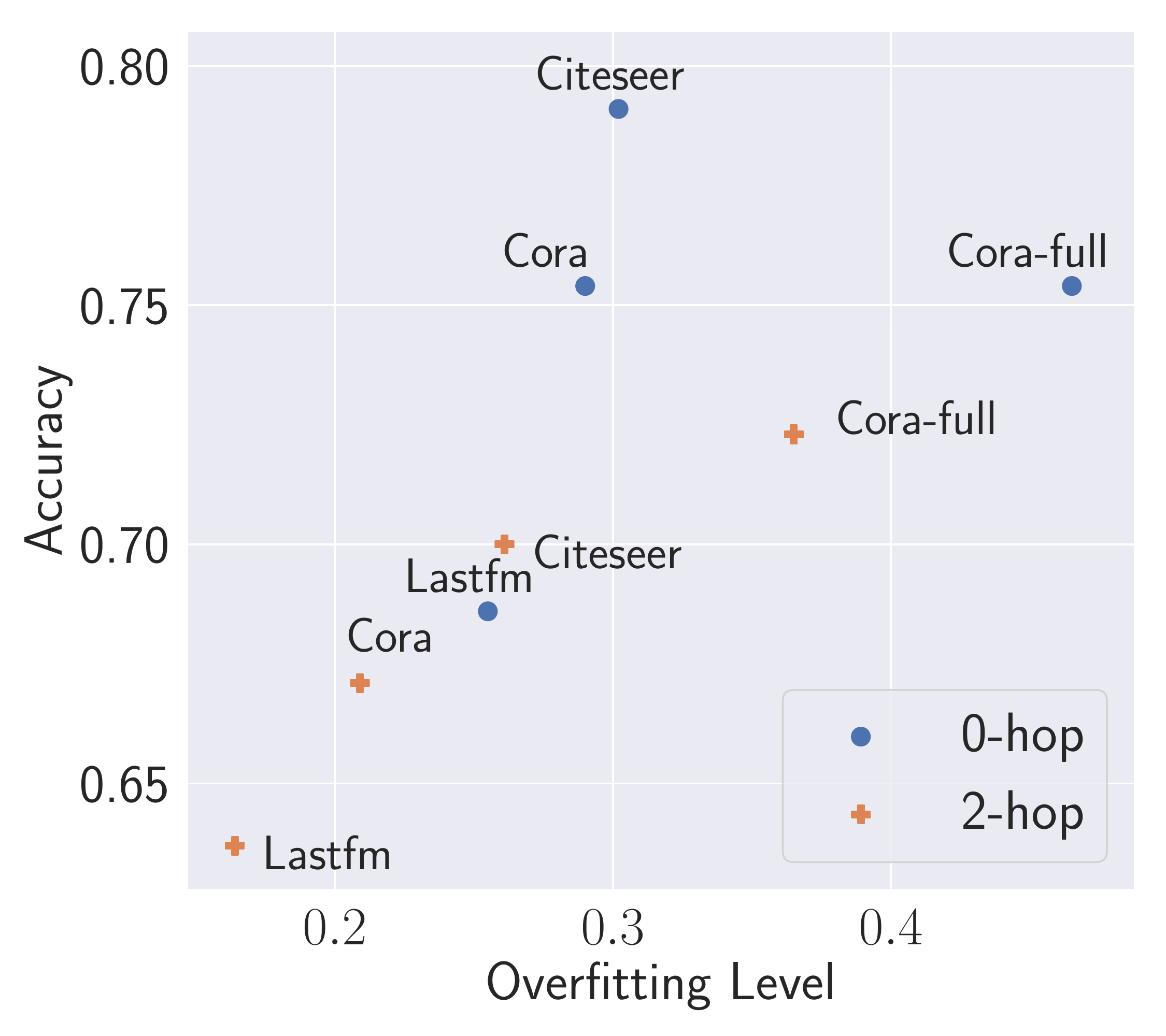}
\caption{GraphSAGE}
\label{figure:overfitting_graphsage}
\end{subfigure}
\begin{subfigure}{0.66\columnwidth}
\includegraphics[width=\columnwidth]{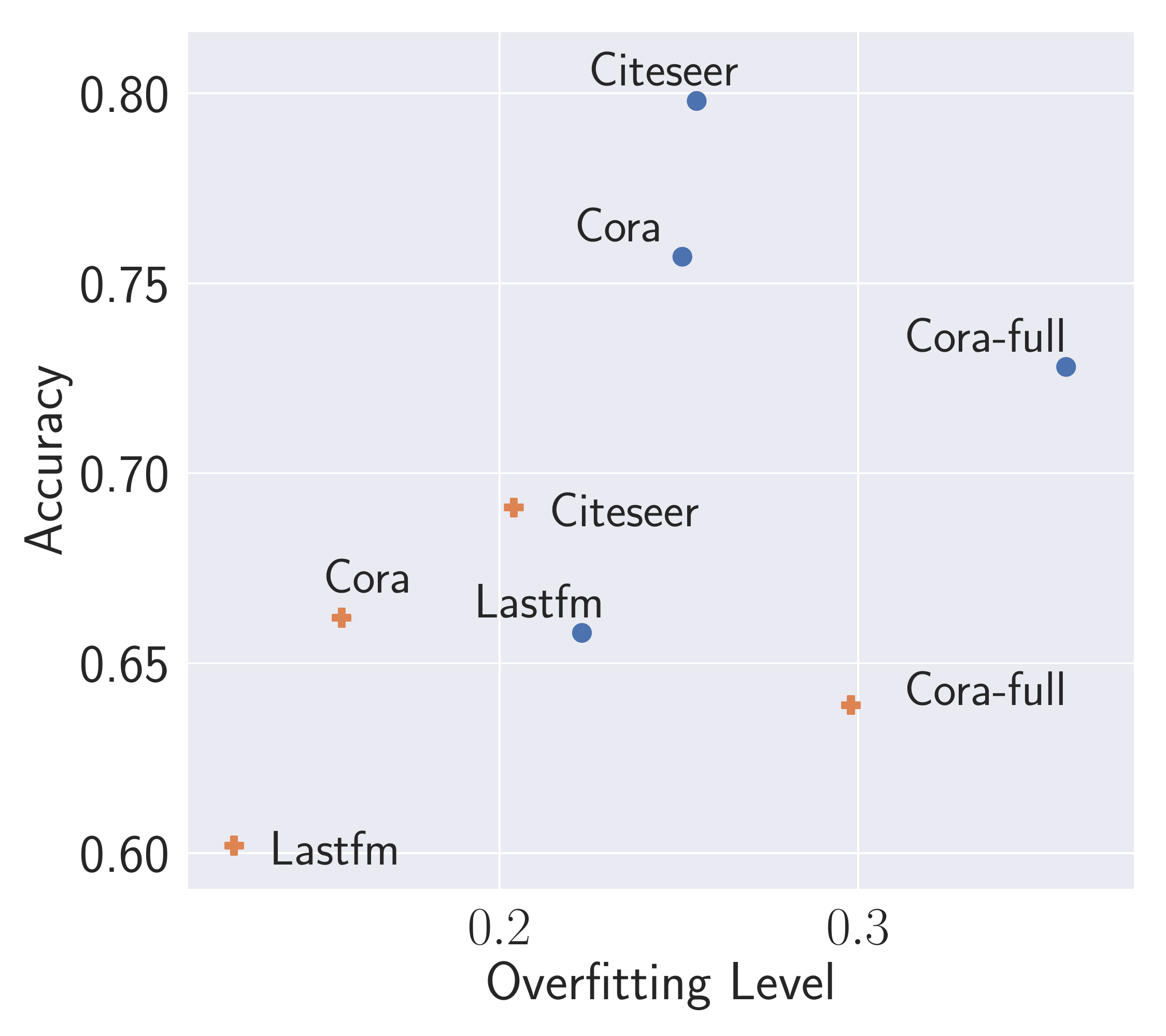}
\caption{GAT}
\label{figure:overfitting_gat}
\end{subfigure}
\begin{subfigure}{0.66\columnwidth}
\includegraphics[width=\columnwidth]{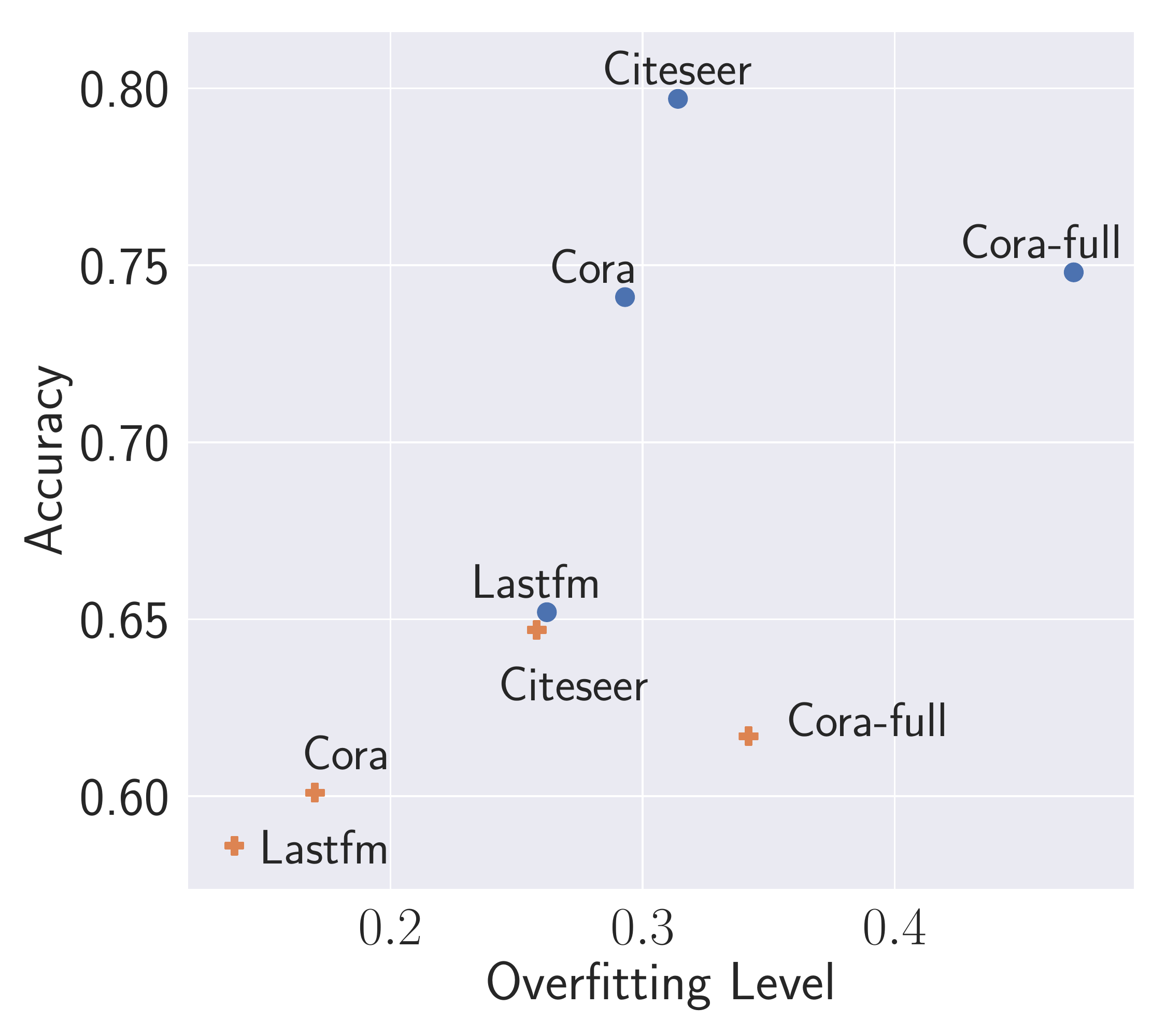}
\caption{GIN}
\label{figure:overfitting_gin}
\end{subfigure}
\caption{The performance of 0-hop and 2-hop attacks for different GNN architectures on four different datasets under different overfitting levels.
The x-axis represents different overfitting levels.
The y-axis represents membership inference attacks' accuracy.}
\label{figure:overfitting}
\end{figure*}

% ----------------------------------------------------
\subsection{0-hop and 2-hop Attacks}
\label{subsection:EvaluationBaiscAttacks}
% ----------------------------------------------------

We first show the membership inference attack performance of the 0-hop and 2-hop attacks in \autoref{table:mia_performance_0hop} and \autoref{table:mia_performance_2hop}, respectively.
We find that compared to the 2-hop attack, the 0-hop attack achieves higher membership inference accuracy.
For instance, the 0-hop attack on GraphSAGE trained on Cora achieves 0.754 accuracy while the accuracy of the corresponding 2-hop attack is only 0.671.
Such observations reveal that a node's 2-hop query to the target GNN leaks less membership information of the node.
This observation is rather interesting since we show that the 2-hop query leads to better node classification accuracy in the original task (see ~\autoref{sec:target_model_performance}).

\begin{table}[!t]
\caption{The performance of 0-hop attacks for different GNN architectures on four different datasets.}
\label{table:mia_performance_0hop}
\centering
\renewcommand{\arraystretch}{1.1}
\begin{tabular}{c|c c c}
\toprule
& \multicolumn{3}{c}{Target Model}\\
Dataset  & GraphSAGE & GIN & GAT \\
\midrule
Cora & 0.754 & 0.741 & 0.757\\
Citeseer & 0.791 & 0.797 & 0.798\\
Cora-full & 0.754 & 0.748 & 0.728\\
Lastfm & 0.686 & 0.652 & 0.658\\
\bottomrule
\end{tabular}
\end{table}

\begin{table}[!t]
\caption{The performance of 2-hop attacks for different GNN architectures on four different datasets.}
\label{table:mia_performance_2hop}
\centering
\renewcommand{\arraystretch}{1.1}
\begin{tabular}{c|c c c}
\toprule
& \multicolumn{3}{c}{Target Model}\\
Dataset  & GraphSAGE & GIN & GAT \\
\midrule
Cora & 0.671 & 0.601 & 0.662\\
Citeseer & 0.700 & 0.647 & 0.691\\
Cora-full & 0.723 & 0.617 & 0.639\\
Lastfm & 0.637 & 0.586 & 0.602\\
\bottomrule
\end{tabular}
\end{table}

To investigate the reason behind this, we visualize the ratio of true positive (TP), false positive (FP), true negative (TN), and false negative (FN) nodes for the 0-hop and 2-hop attacks in \autoref{figure:tpfp_analysis}.
We observe that both attacks achieve a similar true positive rate.
It is reasonable since if a target node is a member, then the target model gives a relatively confident prediction for both its 0-hop and 2-hop queries, and this confident prediction is exploited by both attack models to distinguish the node from non-members.
Meanwhile, the 2-hop attack has a higher ratio of FP (misclassifying non-members as members) than the 0-hop attack.
One reason might be a non-member node's 2-hop subgraph may contain some member nodes.
When the attack model makes a prediction for the non-member node with its 2-hop query, it aggregates the information from the member nodes that might exist in its 2-hop subgraph, thus yields a less accurate prediction.

Similar to previous work~\cite{SSSS17,SZHBFB19}, we measure the relationship between overfitting and attack performance.
The overfitting level is quantified by the difference between training accuracy and testing accuracy of the target model.
In \autoref{figure:overfitting}, we observe that the attack performance is strongly correlated with the overfitting level.
Specifically, in \autoref{figure:overfitting_graphsage}, for the 2-hop attack, the overfitting level for GraphSAGE on the Lastfm dataset is 0.164 and the attack accuracy is 0.637, while a higher overfitting level (0.261) and attack accuracy (0.700) can be observed on the Citeseer dataset.
Also, compared to 0-hop query, 2-hop query has a lower overfitting level, this is due to the fact that 2-hop query achieves better testing accuracy (see \autoref{figure:target_model_performance}).

\begin{figure*}[!t]
\centering
\begin{subfigure}{0.50\columnwidth}
\includegraphics[width=\columnwidth]{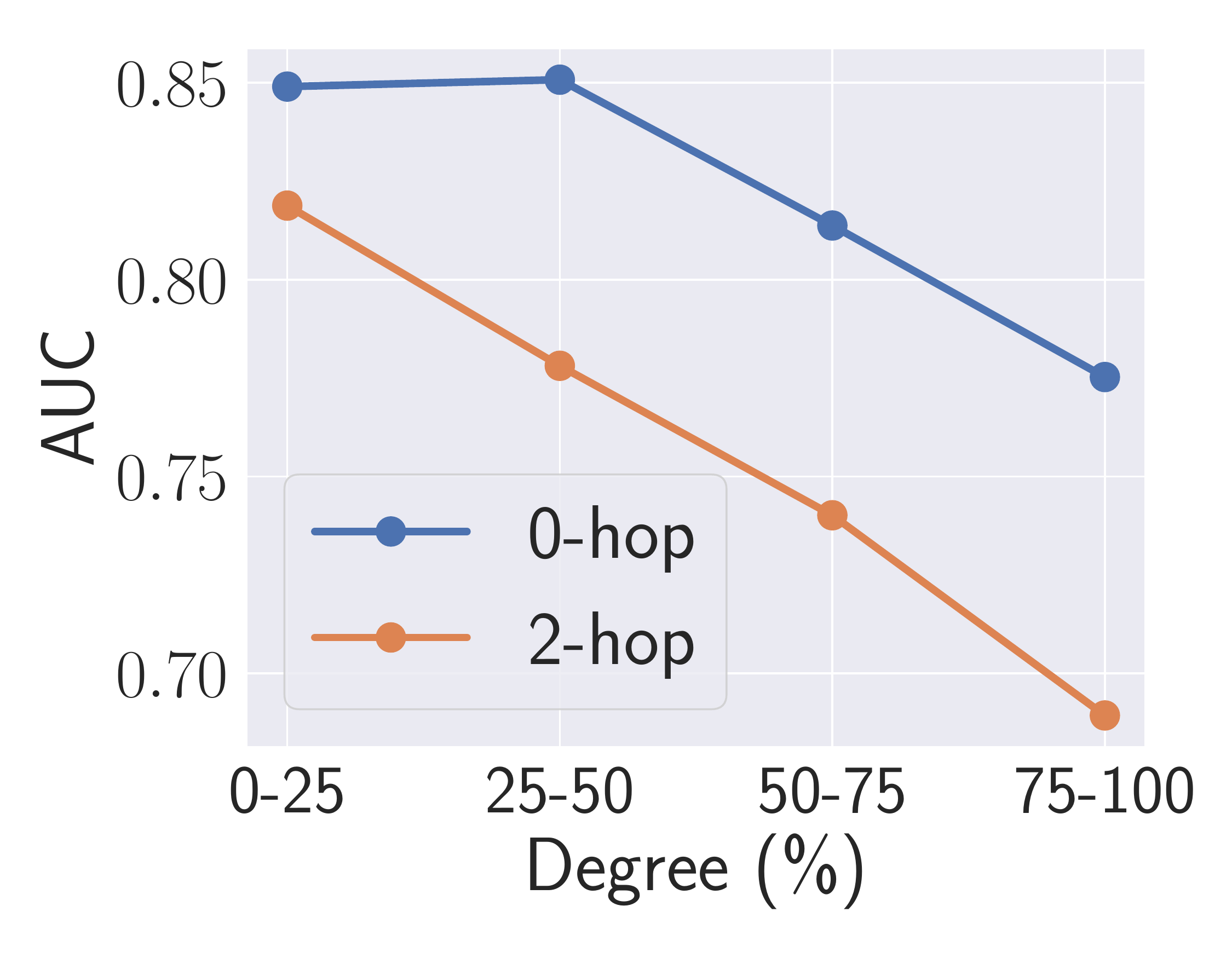}
\caption{Cora}
\label{figure:degree_auc_Cora}
\end{subfigure}
\begin{subfigure}{0.50\columnwidth}
\includegraphics[width=\columnwidth]{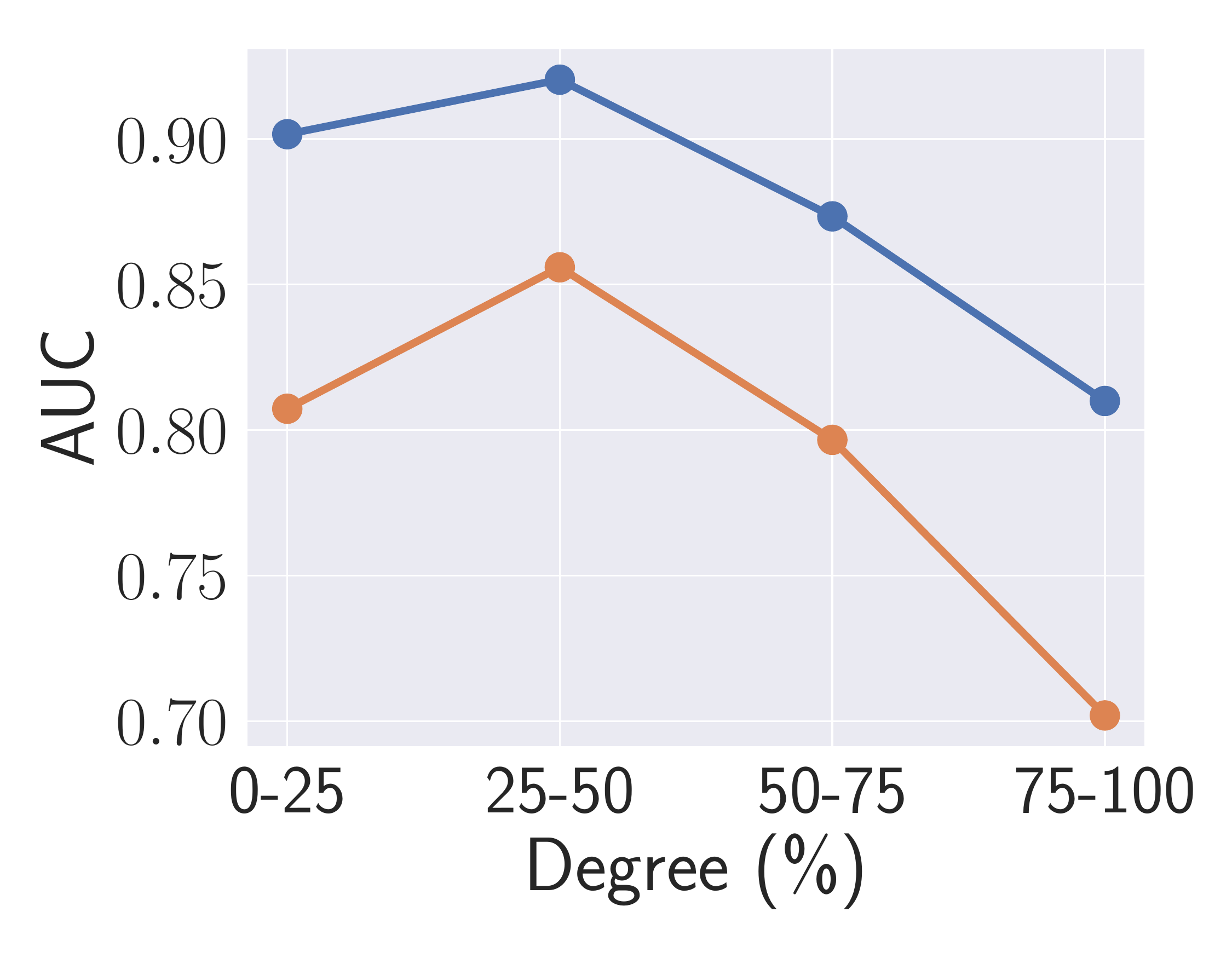}
\caption{Citeseer}
\label{figure:degree_auc_Citeseer}
\end{subfigure}
\begin{subfigure}{0.50\columnwidth}
\includegraphics[width=\columnwidth]{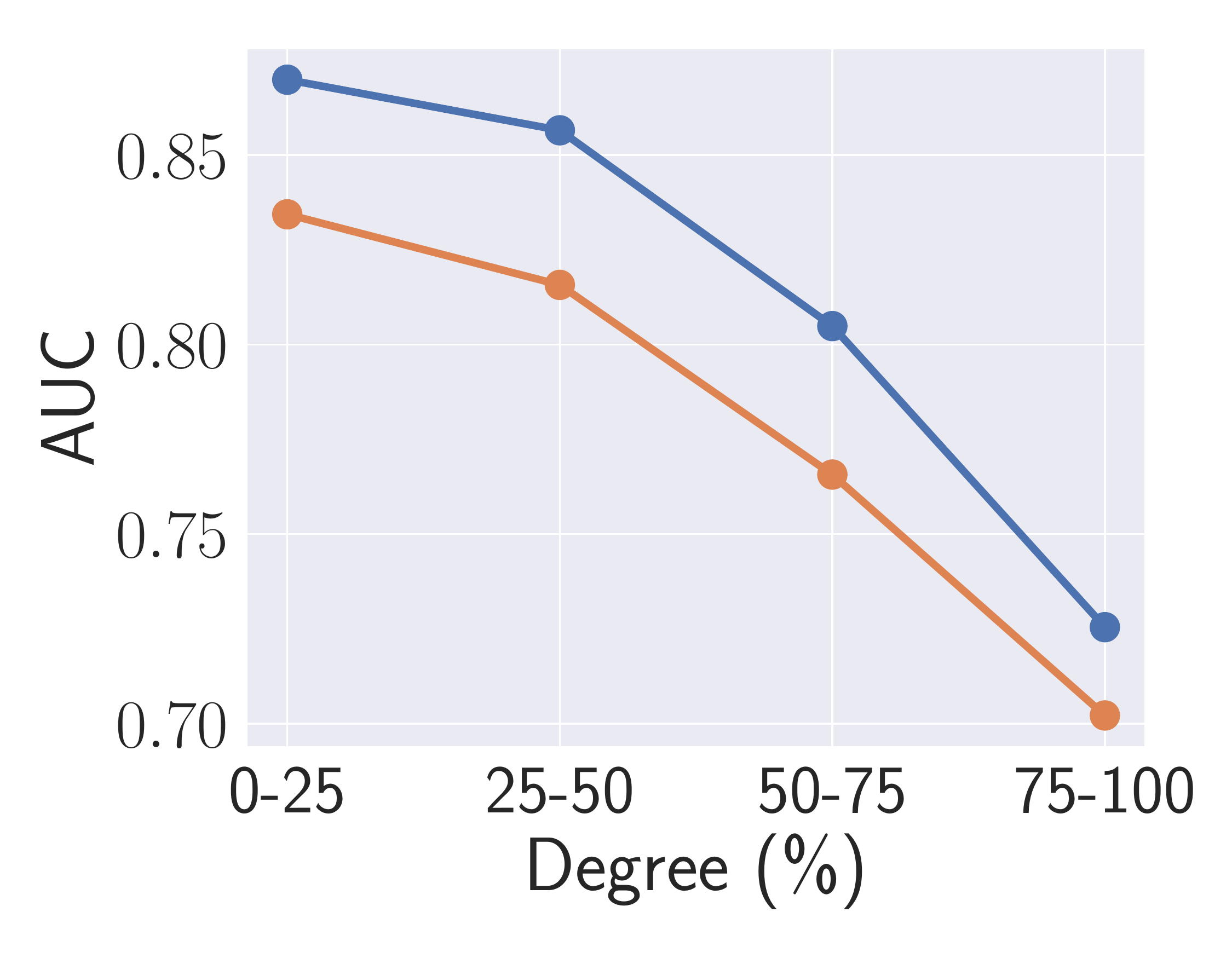}
\caption{Cora-full}
\label{figure:degree_auc_Corafull}
\end{subfigure}
\begin{subfigure}{0.50\columnwidth}
\includegraphics[width=\columnwidth]{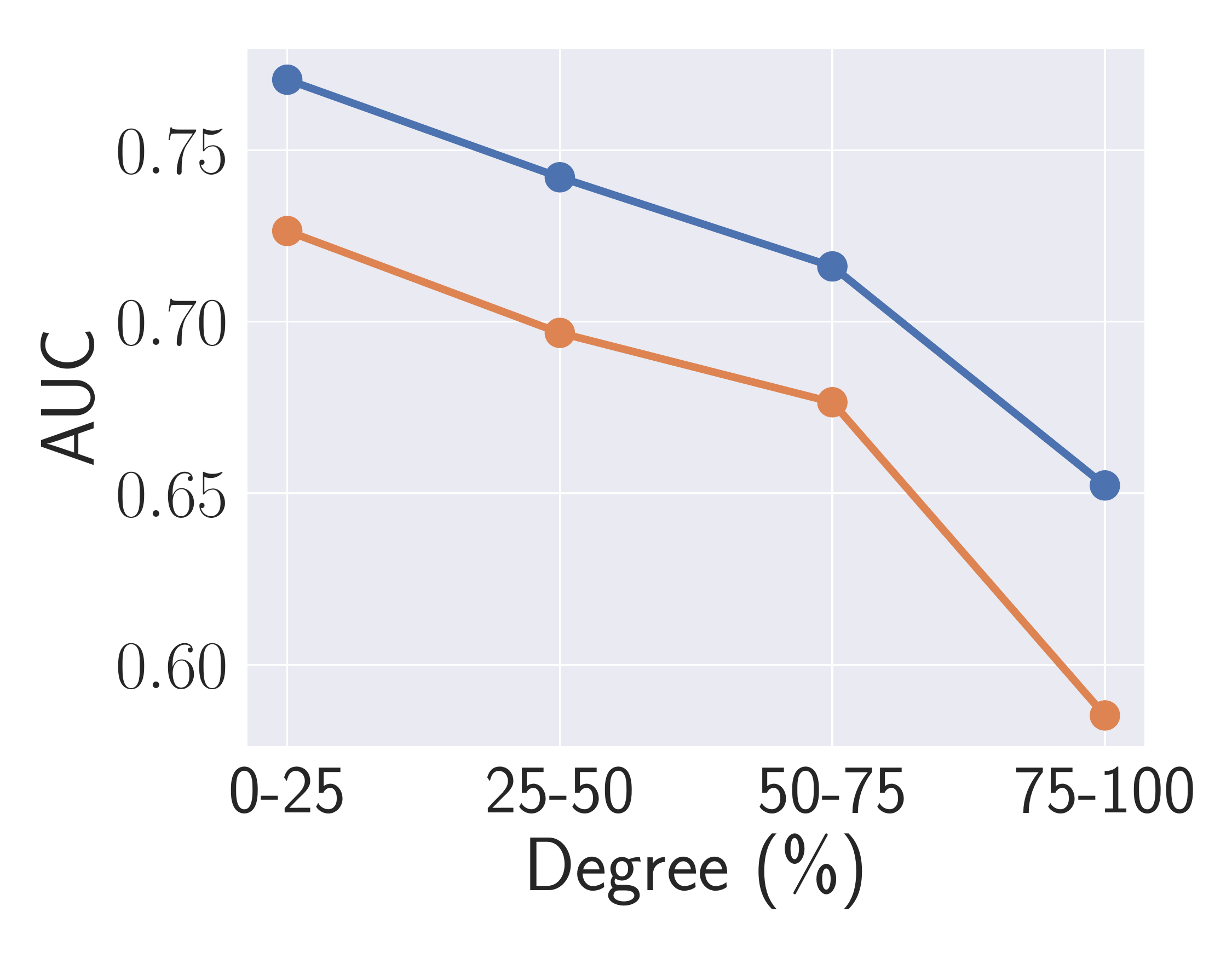}
\caption{Lastfm}
\label{figure:degree_auc_Lastfm}
\end{subfigure}
\caption{AUC for 0-hop and 2-hop attacks on different groups of nodes categorized by degree on four different datasets. 
The architecture of both target and shadow model is GraphSAGE.
The x-axis represents different groups, e.g, 0-25 means the group of nodes whose degrees are in the lowest 25\% of the dataset.
The y-axis represents the AUC.}
\label{figure:micro_degree_vs_auc}
\end{figure*}

\mypara{Node Property}
We next investigate which kinds of nodes are more prone to membership inference.
To this end, we calculate three metrics for each node, i.e., degree, ego density, and feature similarity.
The first two are related to a node's graph property and the last one focuses on the node's feature.

\begin{itemize}
\item \textbf{Degree.}
For a given node $\TargetNode$, the degree of the node is defined as the number of edges connected to it.
\item \textbf{Ego Density.}
Ego density measures the graph density of a node $\TargetNode$'s 2-hop subgraph $\Subgraph^{2}(\TargetNode)$.
\item \textbf{Feature Similarity.}
Feature similarity measures how similar a node $\TargetNode$'s feature to nodes' features in its 2-hop subgraph $\Subgraph^{2}(\TargetNode)$.
Specifically, we calculate the similarity (cosine similarity) between the feature of $\TargetNode$ and the feature of each node in $\Subgraph^{2}(\TargetNode)$.
Then, we average all the similarity.
\end{itemize}

\begin{figure*}[!t]
\centering
\begin{subfigure}{0.50\columnwidth}
\includegraphics[width=\columnwidth]{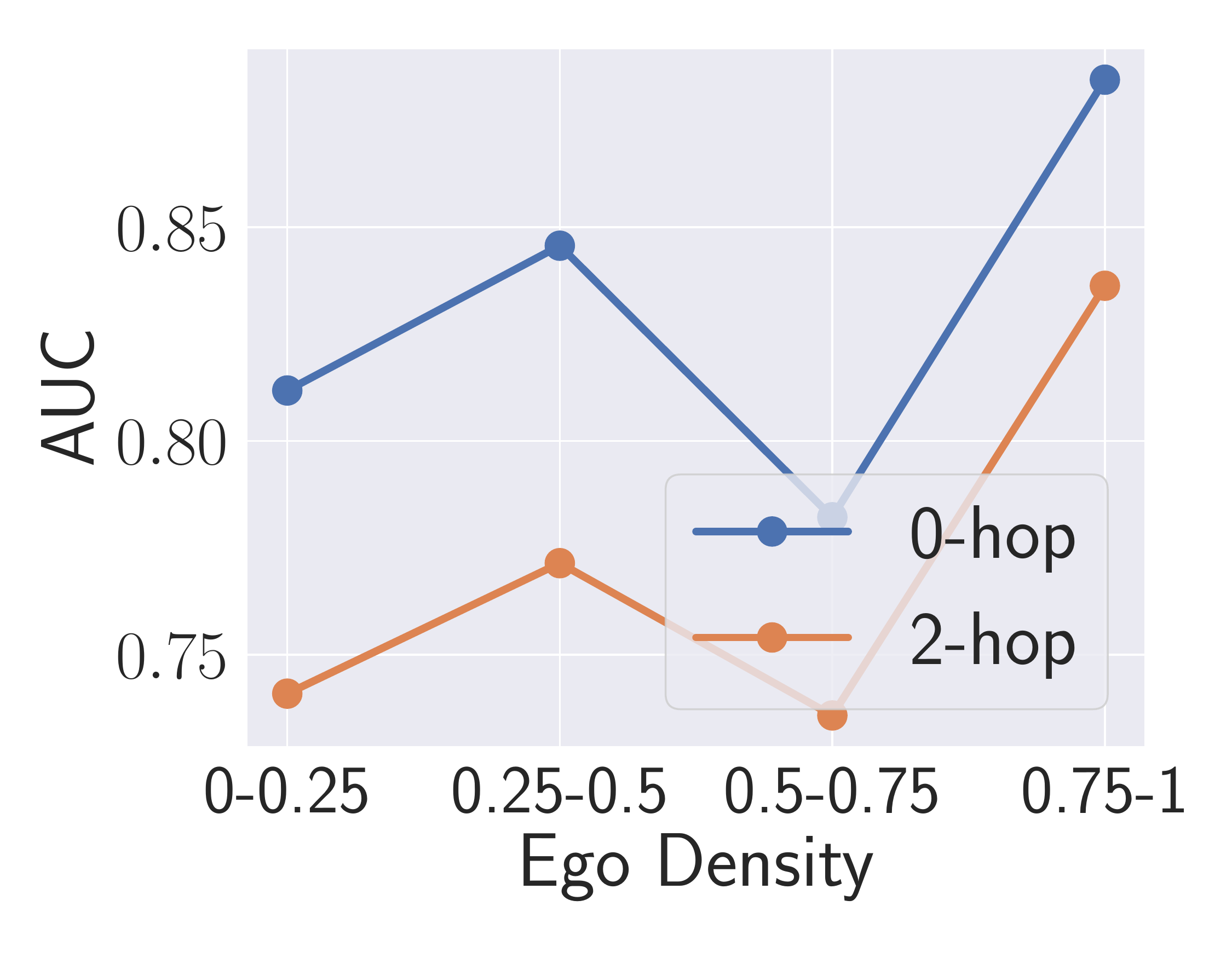}
\caption{Cora}
\label{figure:ego_density_auc_Cora}
\end{subfigure}
\begin{subfigure}{0.50\columnwidth}
\includegraphics[width=\columnwidth]{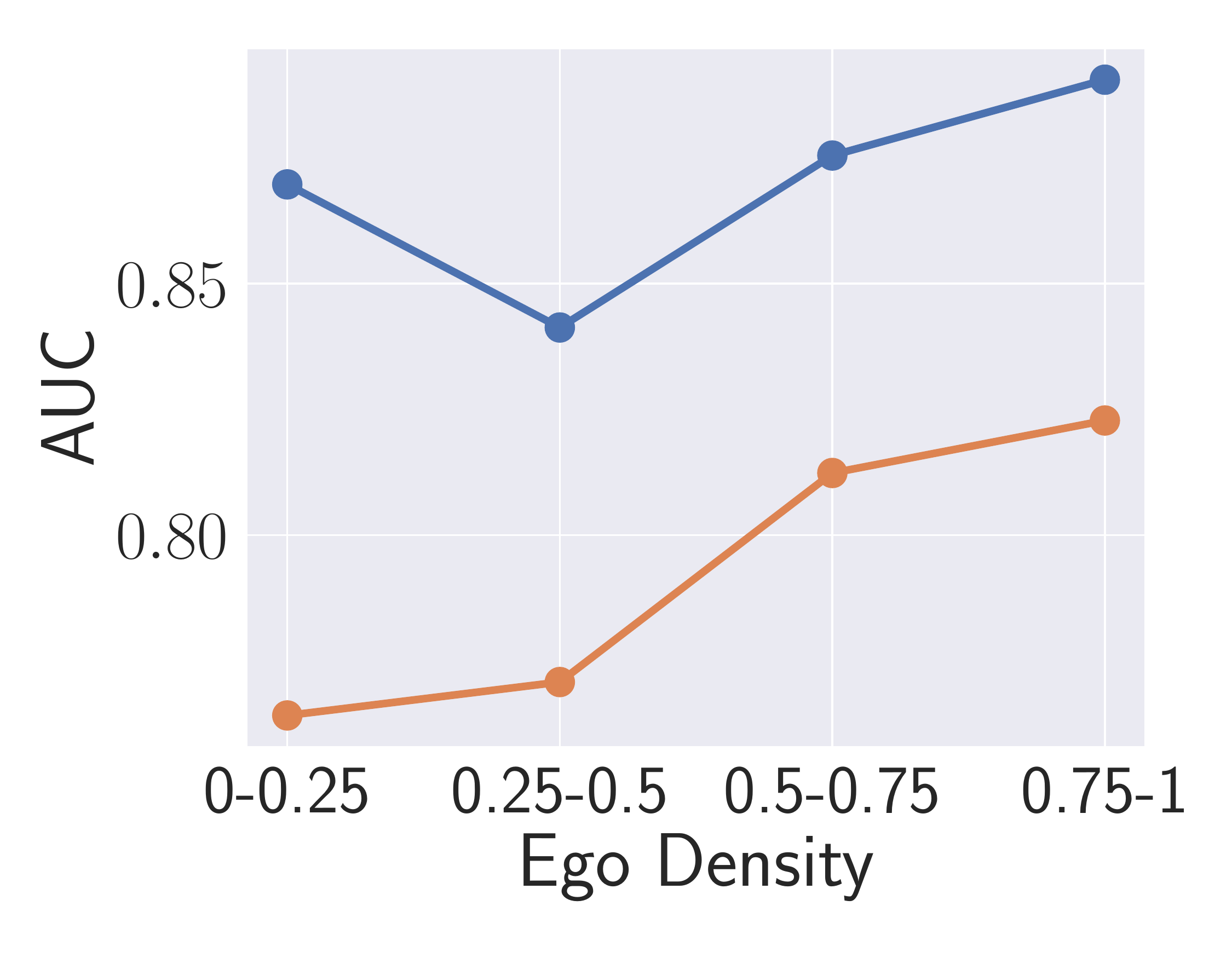}
\caption{Citeseer}
\label{figure:ego_density_auc_Citeseer}
\end{subfigure}
\begin{subfigure}{0.50\columnwidth}
\includegraphics[width=\columnwidth]{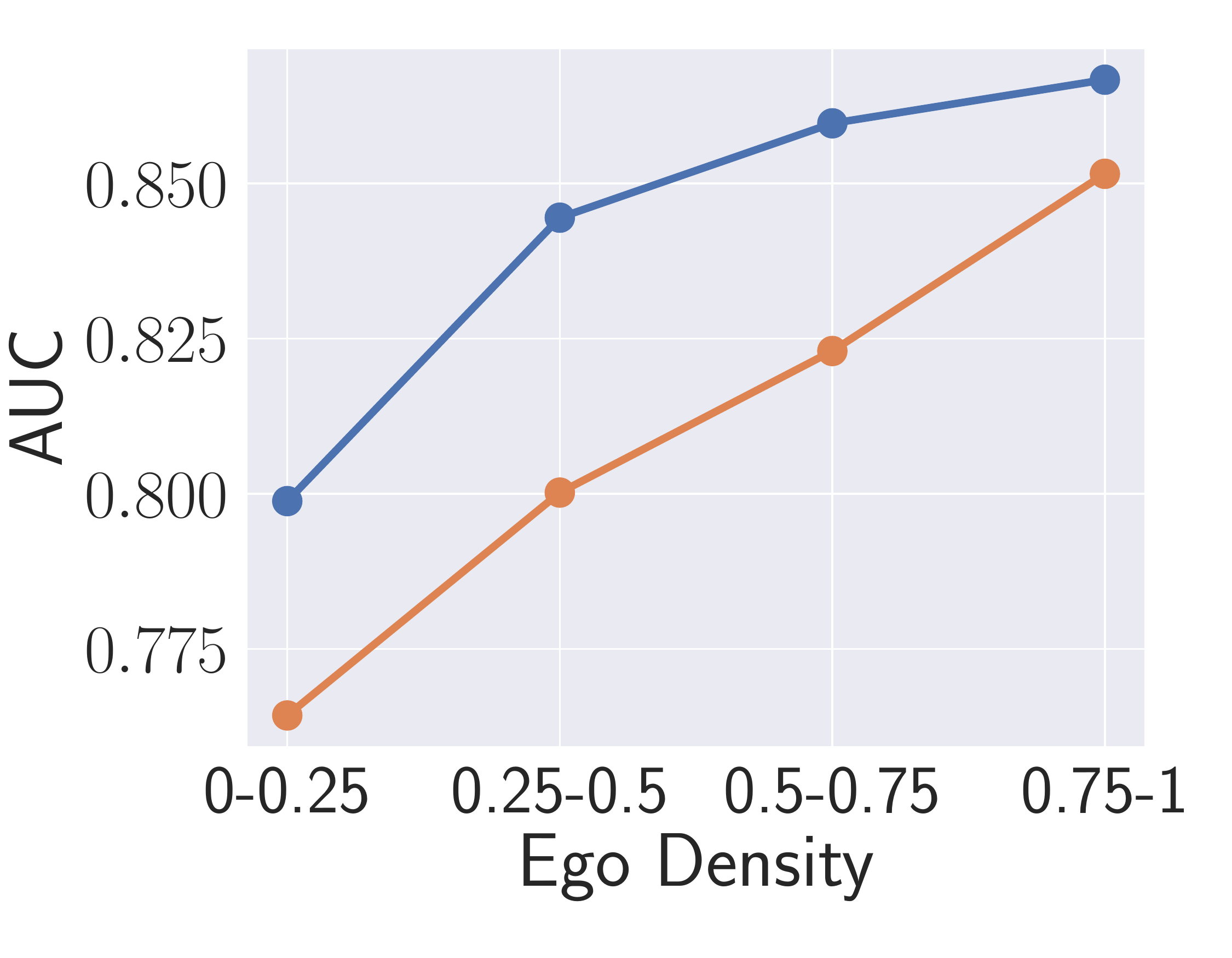}
\caption{Cora-full}
\label{figure:ego_density_auc_Cora-full}
\end{subfigure}
\begin{subfigure}{0.50\columnwidth}
\includegraphics[width=\columnwidth]{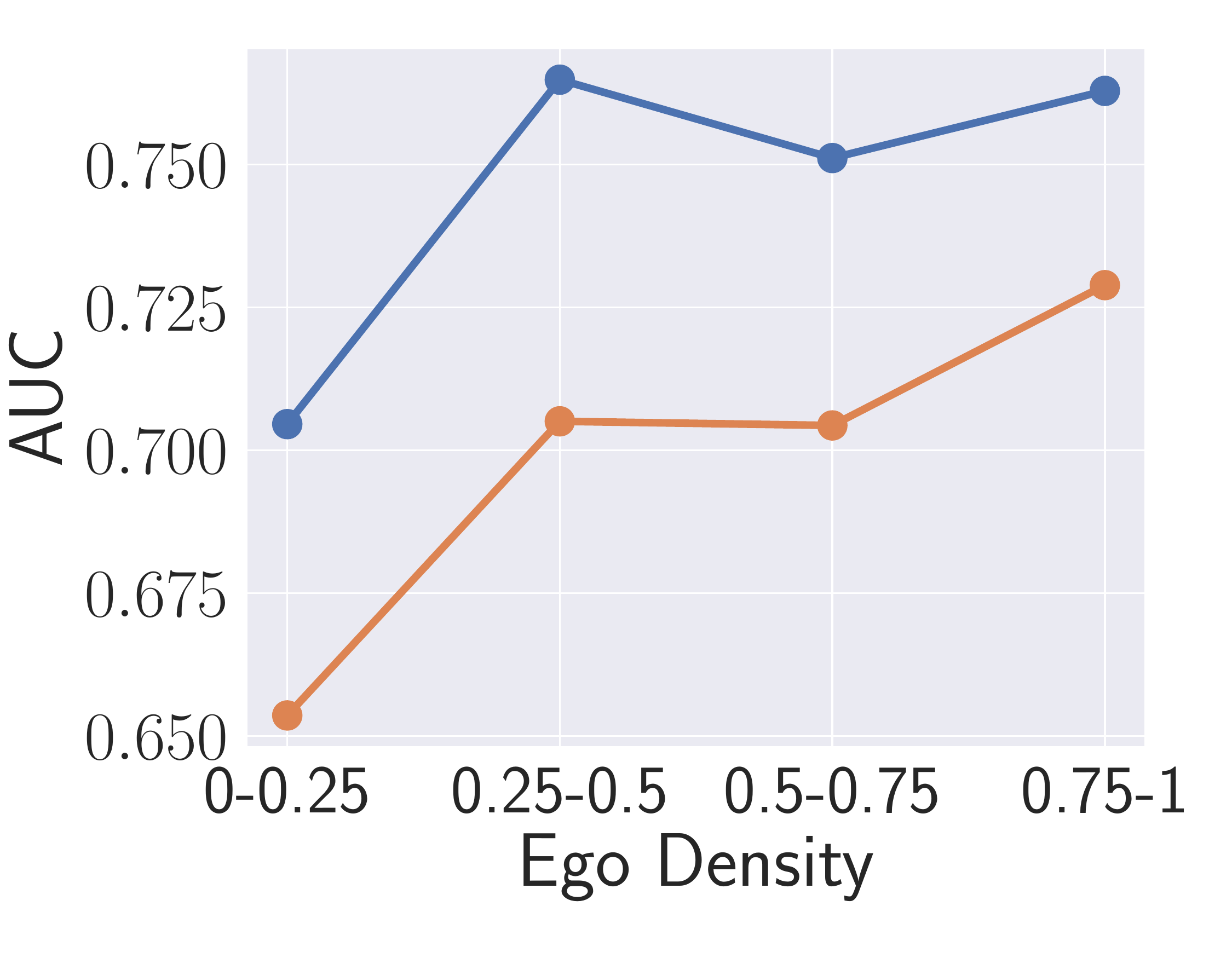}
\caption{Lastfm}
\label{figure:ego_density_auc_Lastfm}
\end{subfigure}
\caption{AUC for 0-hop and 2-hop attacks on different groups of nodes categorized by ego density on four different datasets. 
The architecture of both target and shadow model is GraphSAGE.
The x-axis represents different groups, e.g, 0-0.25 means the group of nodes whose ego density values are in the range of 0 and 0.25.
The y-axis represents the AUC.}
\label{figure:micro_density_vs_auc}
\end{figure*}

\begin{figure*}[!t]
\centering
\begin{subfigure}{0.50\columnwidth}
\includegraphics[width=\columnwidth]{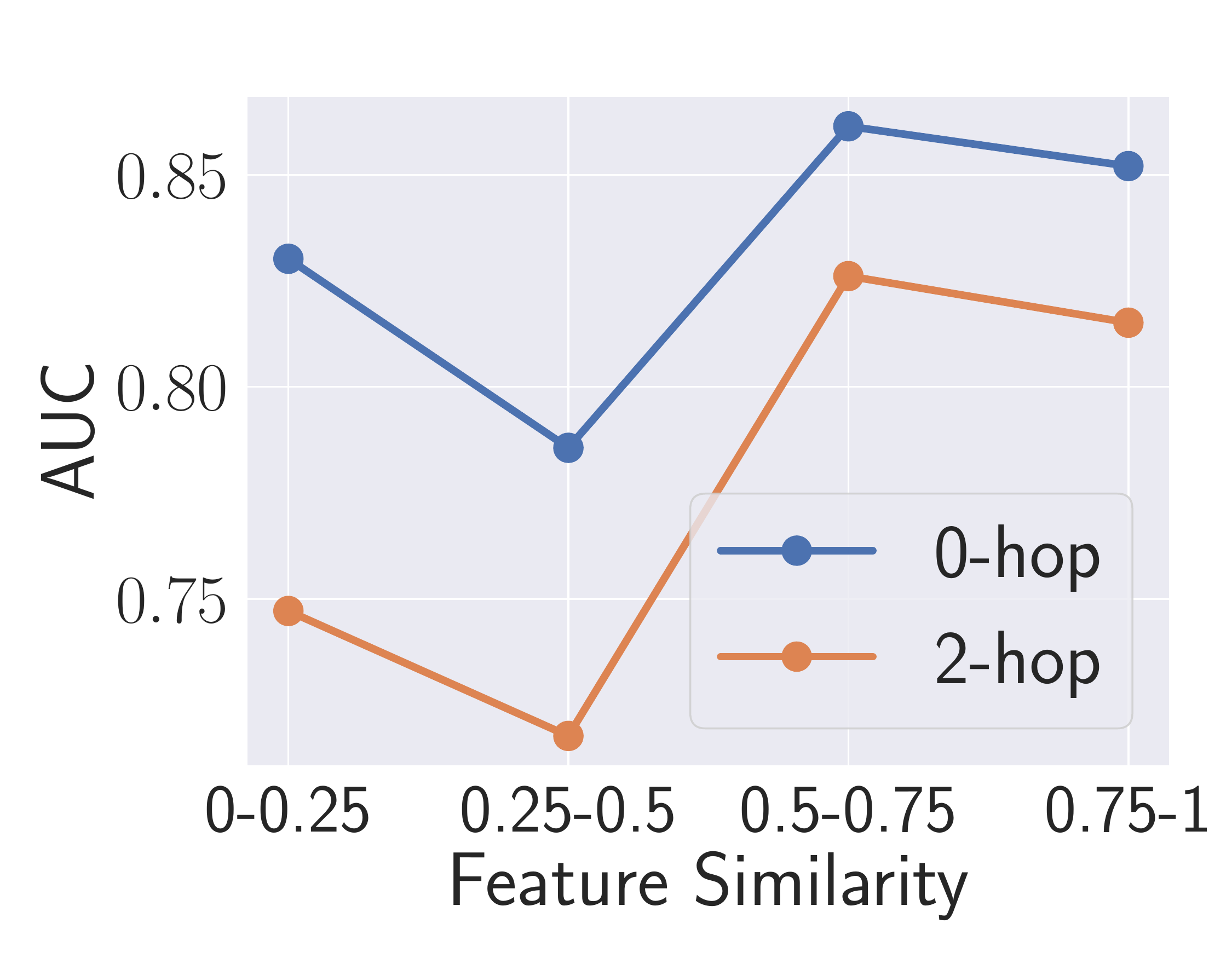}
\caption{Cora}
\label{figure:feature_similarity_auc_Cora}
\end{subfigure}
\begin{subfigure}{0.50\columnwidth}
\includegraphics[width=\columnwidth]{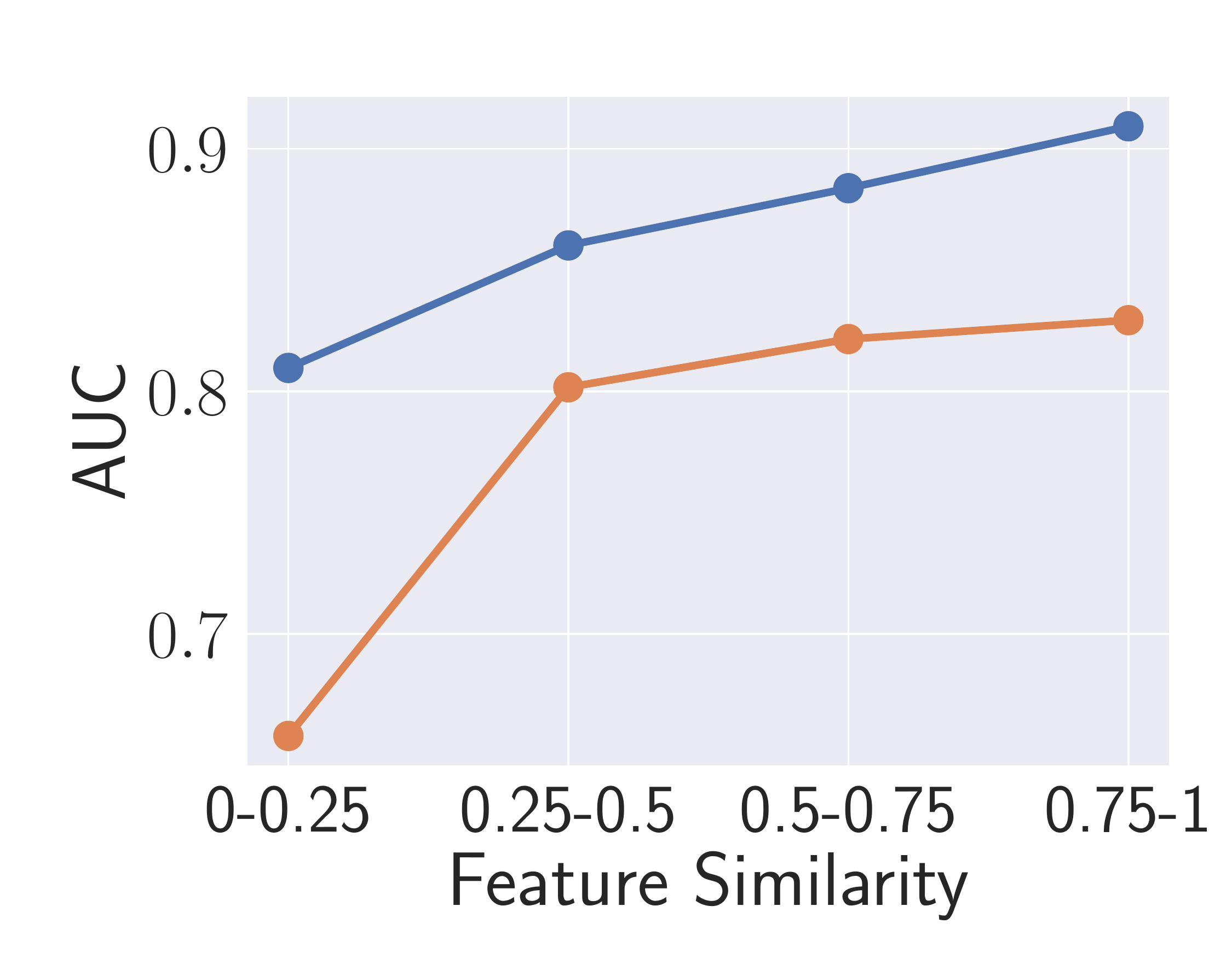}
\caption{Citeseer}
\label{figure:feature_similarity_auc_Citeseer}
\end{subfigure}
\begin{subfigure}{0.50\columnwidth}
\includegraphics[width=\columnwidth]{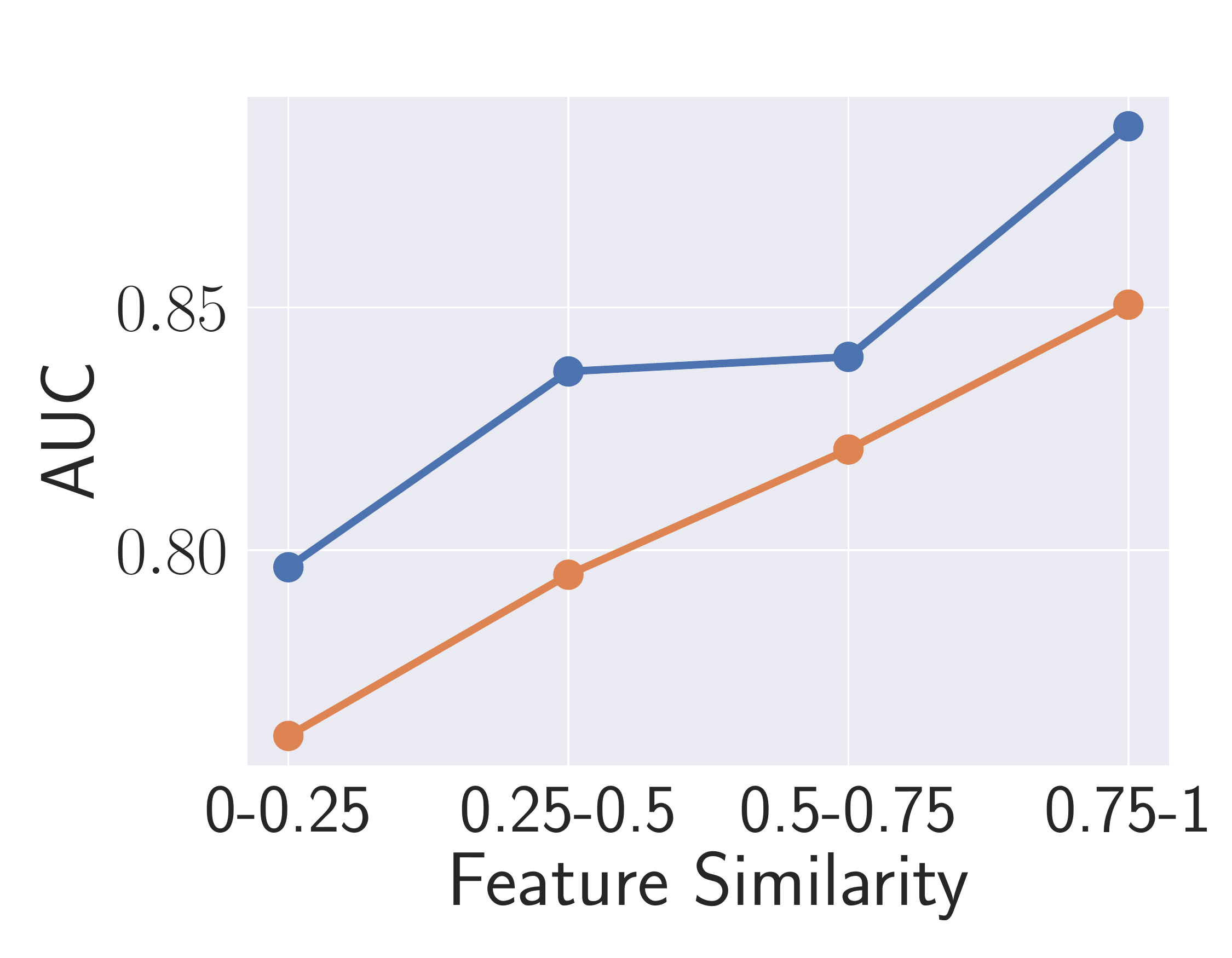}
\caption{Cora-full}
\label{figure:feature_similarity_auc_Cora-full}
\end{subfigure}
\begin{subfigure}{0.50\columnwidth}
\includegraphics[width=\columnwidth]{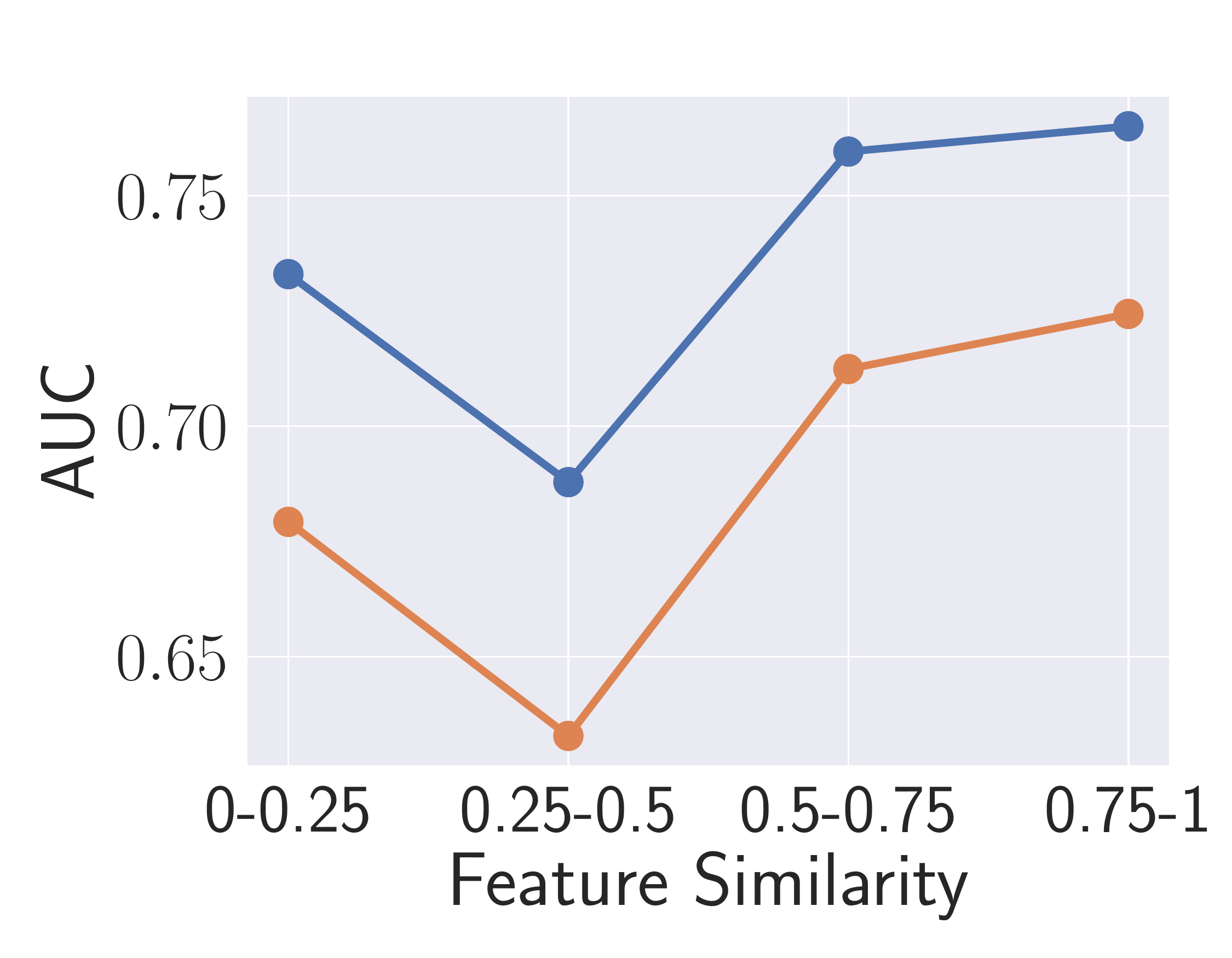}
\caption{Lastfm}
\label{figure:feature_similarity_auc_Lastfm}
\end{subfigure}
\caption{AUC for 0-hop and 2-hop attacks on different groups of nodes categorized by feature similarity on four different datasets.
The architecture of both target and shadow models is GraphSAGE.
The x-axis represents different groups, e.g, 0-0.25 means the group of nodes whose feature similarity values are in the range of 0 and 0.25.
The y-axis represents the AUC.}
\label{figure:micro_similarity_vs_auc}
\end{figure*}

We categorize all the nodes in $\TargetDataset\Train$ and $\TargetDataset\Test$, i.e., the attack model's testing dataset, into four different groups based on their degrees, ego density, and feature similarity, respectively.
The results are summarized in \autoref{figure:micro_degree_vs_auc}, \autoref{figure:micro_density_vs_auc}, and \autoref{figure:micro_similarity_vs_auc}.
Note that the distribution of member and non-member nodes in each group is not uniform, thus we utilize AUC (area under the ROC curve) to measure the attack performance in each group as AUC is not sensitive to imbalanced classes~\cite{BHPZ17,FLJLPR14}.
In general, we find that higher degree leads to lower AUC score for both 0-hop and 2-hop attacks (see \autoref{figure:micro_degree_vs_auc}).
For instance, for GraphSAGE trained on the Cora dataset, the 0-hop attack's AUC is 0.849 on nodes in the lowest 25\% degree group while the AUC is 0.775 in the highest 25\% degree group.
Recall that during the training process, each GNN layer generates a node's representation by aggregating its neighbor nodes' representation.
With a higher degree, more neighbor nodes are involved, which may reduce the ``exposure'' of the target node itself, thus lesser membership inference risk. 

In \autoref{figure:micro_density_vs_auc}, we find that larger ego density implies higher attack performance.
For instance, for GraphSAGE trained on Cora-full, the 0-hop attack achieves 0.799 AUC on nodes with less than 0.25 ego density while the AUC increases to 0.867 for nodes with larger than 0.75 ego density.
The reason behind this can be credited to the aggregation function of GNN models. 
Higher density enables a node to participate more times in the aggregation process during training, which results in the model memorizing more information about the node.
Also, if the density of a node's 2-hop subgraph is high, then all the nodes in the subgraph are more likely to share similar features as the node, following social homophily theory~\cite{EK10}.
This further amplifies the influence of the node in the model.

We further measure the relation between attack performance and feature similarity (see \autoref{figure:micro_similarity_vs_auc}).
Our finding reveals that membership inference is indeed more effective when the target node has a larger feature similarity with its neighbors.
For GraphSAGE trained on Citeseer, the 2-hop attack's AUC increases from 0.657 to 0.829 when the feature similarity increase from less than 0.25 to larger than 0.75.

% ----------------------------------------------------
\subsection{Combined Attacks}
\label{subsection:EvaluationCombinedAttacks}
% ----------------------------------------------------

\begin{figure*}[!t]
\centering
\begin{subfigure}{0.50\columnwidth}
\includegraphics[width=\columnwidth]{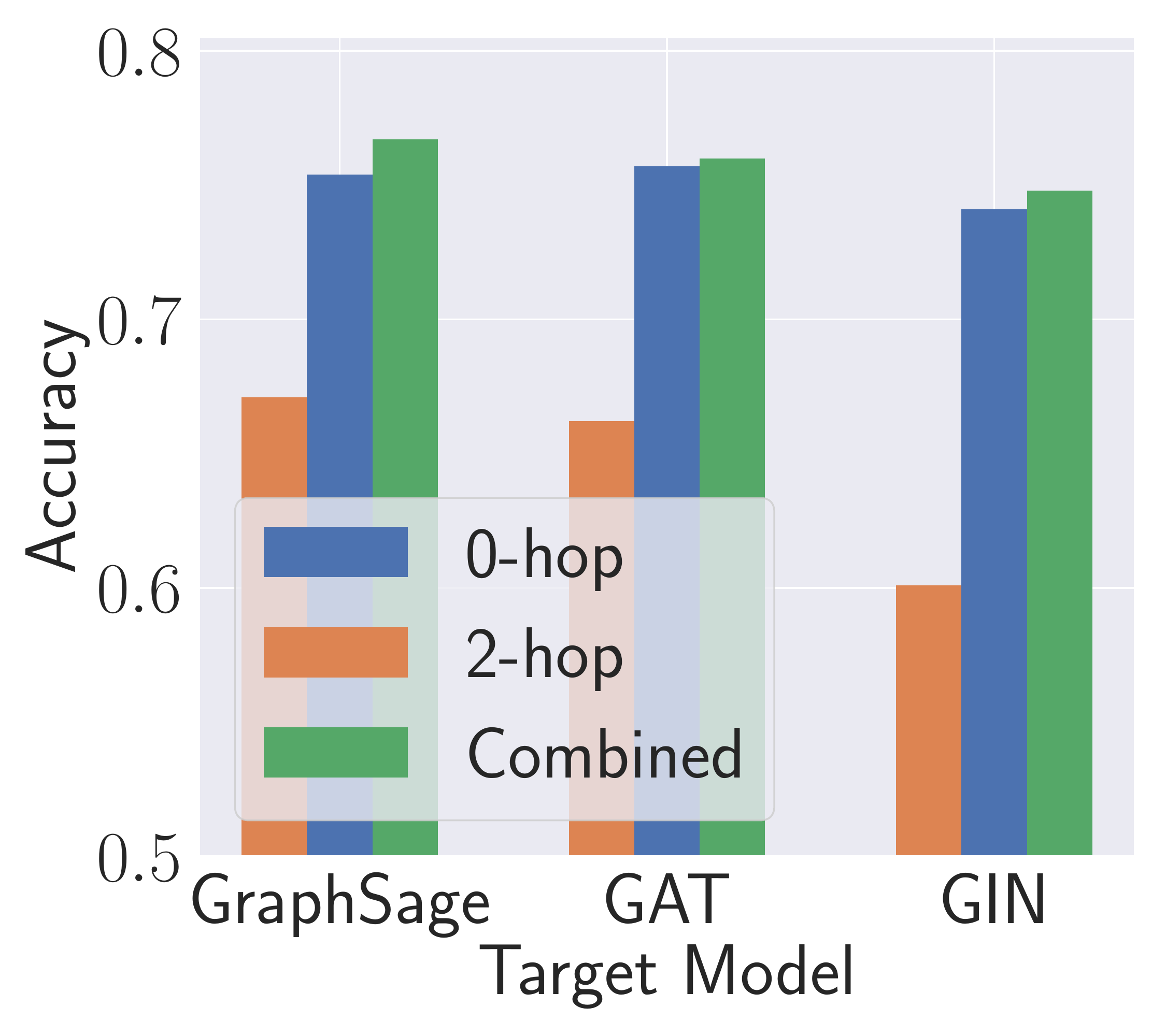}
\caption{Cora}
\label{figure:attack_Cora}
\end{subfigure}
\begin{subfigure}{0.50\columnwidth}
\includegraphics[width=\columnwidth]{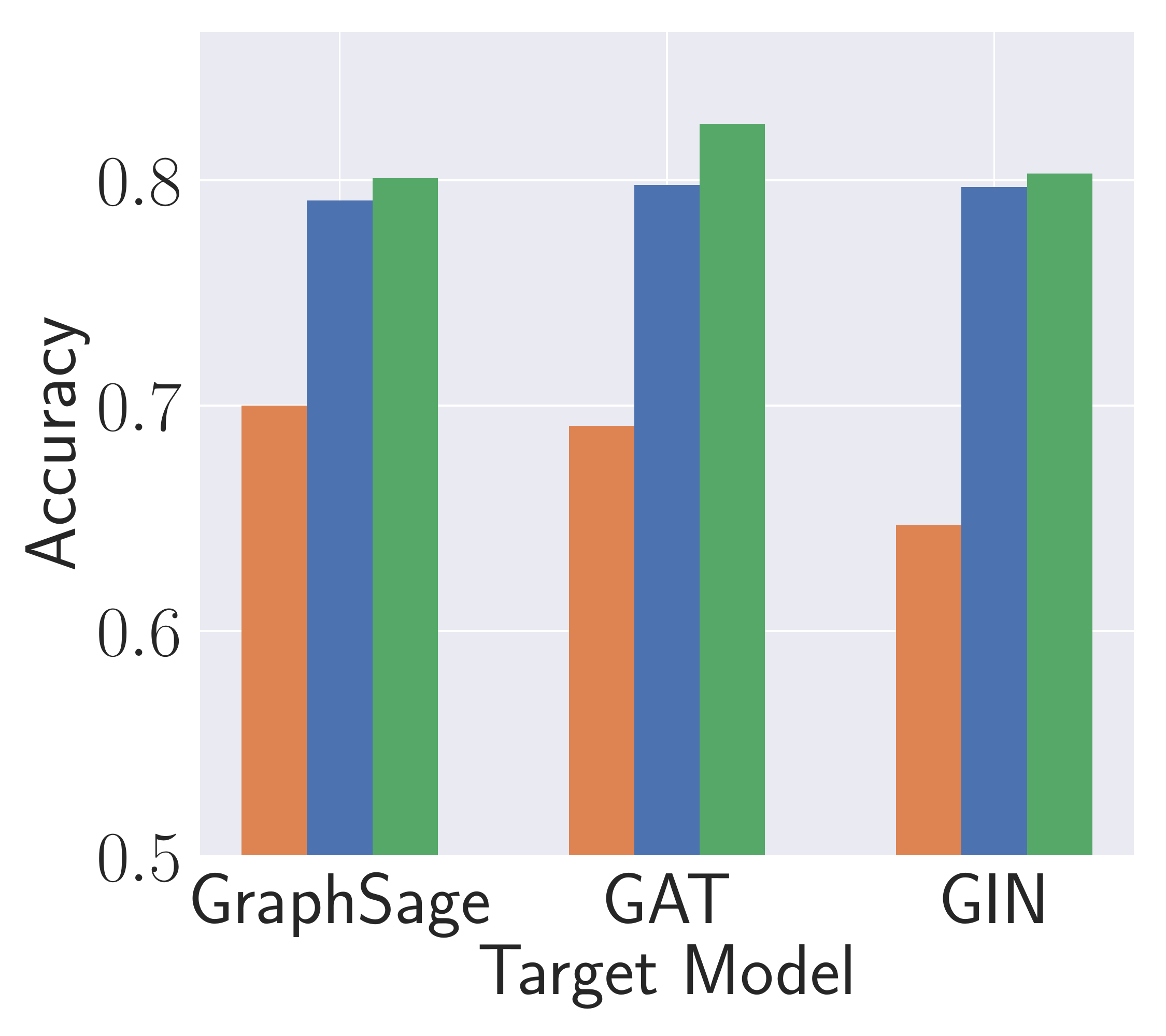}
\caption{Citeseer}
\label{figure:attack_Citeseer}
\end{subfigure}
\begin{subfigure}{0.50\columnwidth}
\includegraphics[width=\columnwidth]{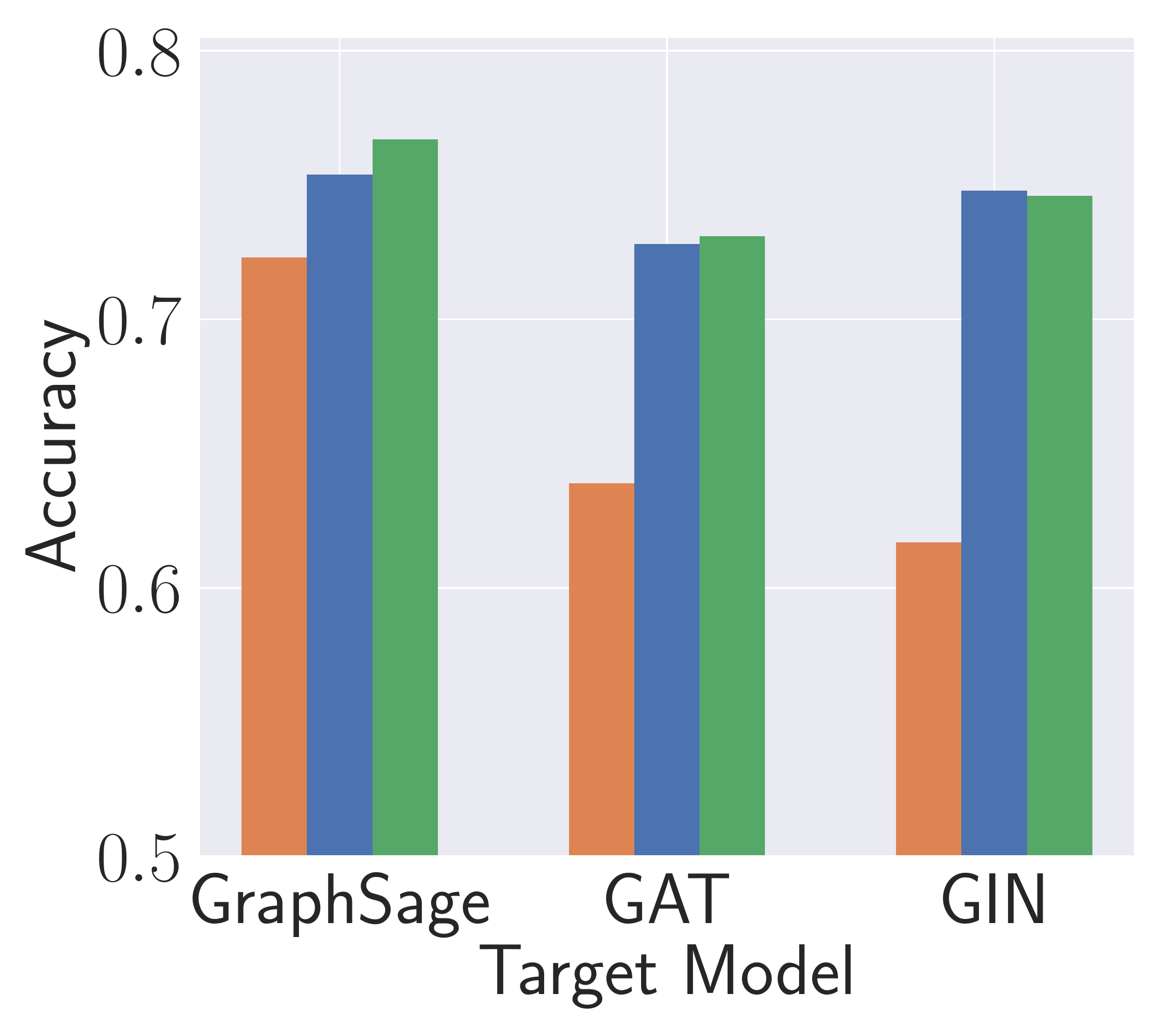}
\caption{Cora-full}
\label{figure:attack_Corafull}
\end{subfigure}
\begin{subfigure}{0.50\columnwidth}
\includegraphics[width=\columnwidth]{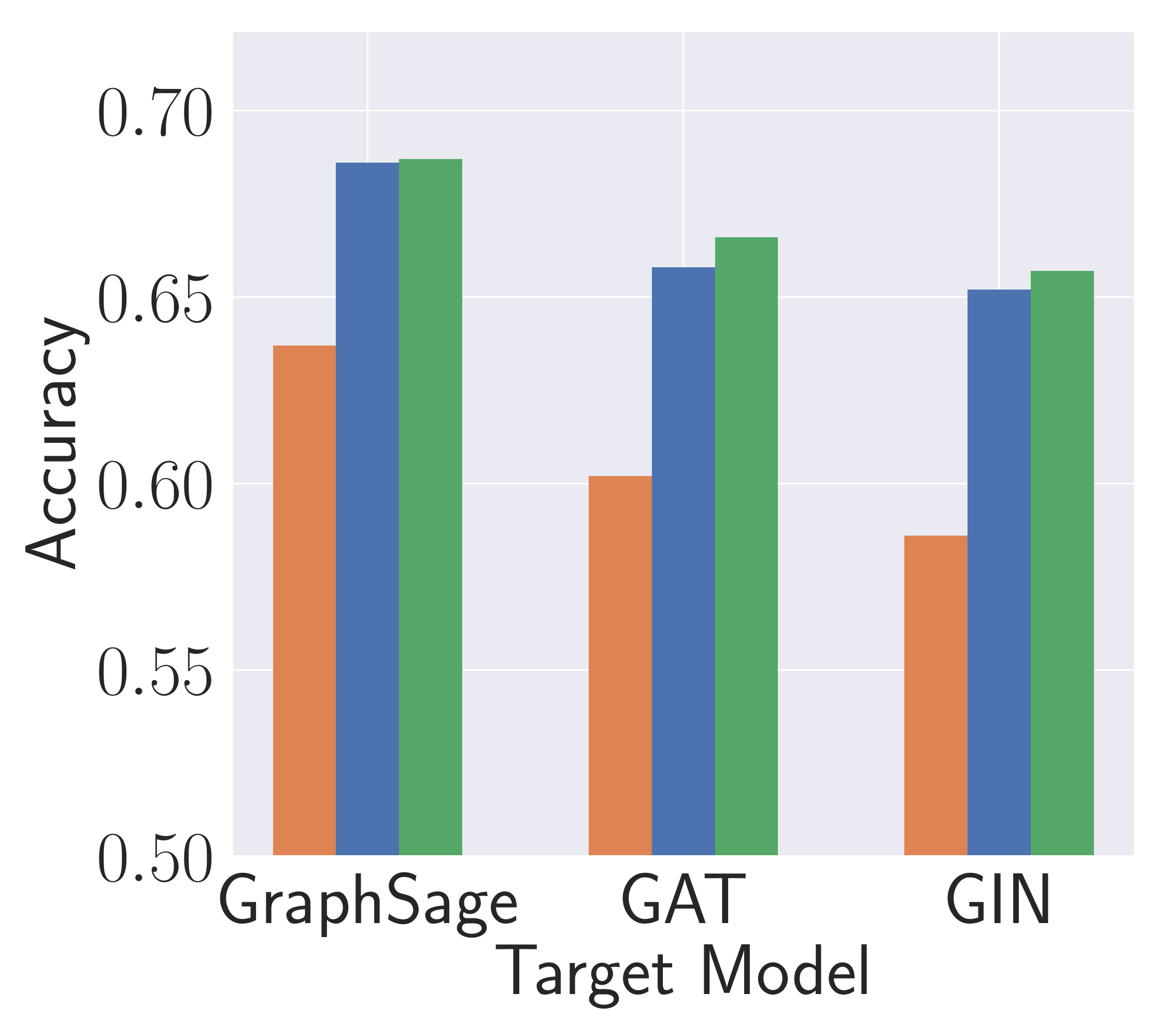}
\caption{Lastfm}
\label{figure:attack_lastfm}
\end{subfigure}
\caption{The performance of membership inference attacks against different target models' architectures on four different datasets.
The x-axis represents different target models' architectures.
The y-axis represents membership inference attacks' accuracy.}
\label{figure:attack_performance}
\end{figure*}

\begin{figure*}[!t]
\centering
\begin{subfigure}{2\columnwidth}
\includegraphics[width=\columnwidth]{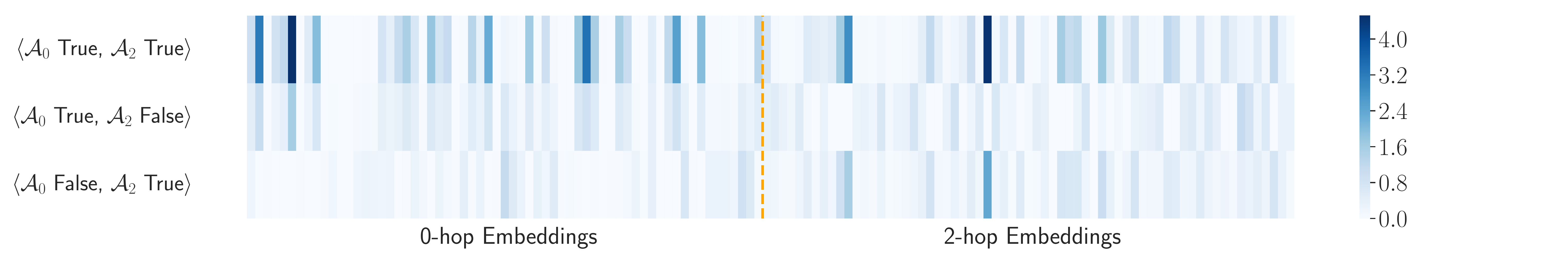}
\caption{Non-members}
\label{figure:embedding_nonmem_Citeseer}
\end{subfigure}
\begin{subfigure}{2\columnwidth}
\includegraphics[width=\columnwidth]{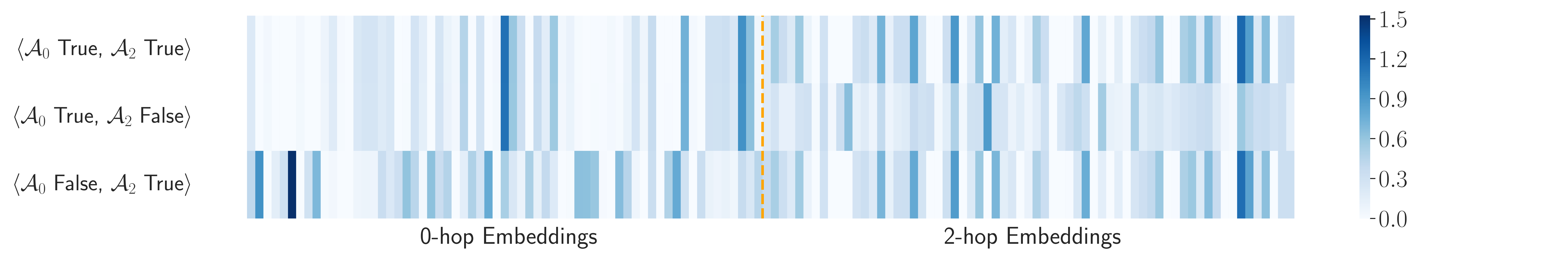}
\caption{Members}
\label{figure:embedding_mem_Citeseer}
\end{subfigure}
\caption{The average embeddings of non-member and member nodes obtained from the combined attack model's hidden layer.
Both the target and shadow models are GraphSAGE trained on Citeseer. 
The left (right) parts are the embeddings generated by the input of the 0-hop (2-hop) attack.}
\label{figure:embedding_combine_attack}
\end{figure*}

The combined attack takes both the inputs of the 0-hop and 2-hop attack models as its input.
Its performance is summarized in \autoref{figure:attack_performance}.
In most cases, we observe that the combined attack reaches the highest membership inference accuracy compared to the 0-hop and 2-hop attacks.
For instance, when the target model is GAT trained on Citeseer, the membership inference accuracy is 0.825 for the combined attack, while only 0.798 and 0.691 for the 0-hop and 2-hop attacks, respectively.
\autoref{figure:tpfp_analysis} further shows the ratio of TP, FP, TN, and FN for the combined attack.

We also visualize the embeddings of non-member and member nodes obtained from the combined attack model's hidden layer (when the target model is GraphSAGE trained on Citeseer) in \autoref{figure:embedding_combine_attack}.
The hidden layer embedding is a 128-dimensional vector, where the first 64 dimension corresponds to the input to the 0-hop attack model (referred to as 0-hop embedding), and the second 64 dimension corresponds to the input to the 2-hop attack model (referred to as 2-hop embedding).
We use an orange dash line to separate them for better visualization. 
In \autoref{figure:embedding_nonmem_Citeseer}, the first row represents the average embeddings of the non-member nodes that are correctly predicted by both the 0-hop and 2-hop attacks.
The second (third) row represents the average embeddings of the non-member nodes that are correctly (wrongly) predicted by the 0-hop attack but wrongly (correctly) by the 2-hop attacks.
We can see that the left parts of the first row and the second row (0-hop embedding) are similar, while the right parts are relatively different.
This indicates that for those non-member nodes that are classified correctly by the 0-hop attack but wrongly by the 2-hop attack, the combined attack is able to follow the prediction of the 0-hop attack.
On the other hand, when the non-member nodes are classified correctly by the 2-hop attack but wrongly by the 0-hop attack, the combined attack follows the prediction of the 2-hop attack as the right parts of the third row resemble the right parts of the first row.
We observe similar trends for member nodes (see \autoref{figure:embedding_mem_Citeseer}).
In conclusion, the combined attack has better performance since it takes the advantage of both 0-hop and 2-hop attacks.

% ----------------------------------------------------
\subsection{Relax Assumptions}
\label{subsection:EvaluationRelaxAssumptions}
% ----------------------------------------------------

\begin{table}[!t]
\caption{The performance of combined attacks when using different distribution shadow datasets to train the shadow models. 
Both the target and shadow models' architecture is GraphSAGE.}
\label{table:attack_performance_different_shadow_dataset}
\centering
\renewcommand{\arraystretch}{1.1}
\begin{tabular}{c|c c c c}
\toprule
& \multicolumn{4}{c}{Shadow Dataset}\\
Target Dataset  & Cora & Citeseer & Cora-full & Lastfm \\
\midrule
Cora & 0.767 & 0.775 & 0.715 & 0.743\\
Citeseer & 0.791 & 0.801 & 0.764 & 0.773\\
Cora-full & 0.721 & 0.736 & 0.767 & 0.705\\
Lastfm & 0.696 & 0.693 & 0.643 & 0.687\\
\bottomrule
\end{tabular}
\end{table}

\begin{table}[!t]
\caption{The performance of combined attacks when using different architectures to establish the shadow models.
The target model's architecture is GraphSAGE.
Both the target and shadow training datasets are from the same distribution.}
\label{table:attack_performance_different_shadow_model}
\centering
\renewcommand{\arraystretch}{1.1}
\begin{tabular}{c|c c c}
\toprule
& \multicolumn{3}{c}{Shadow Model}\\
Dataset  & GraphSAGE & GIN & GAT \\
\midrule
Cora & 0.767 & 0.759 & 0.742\\
Citeseer & 0.801 & 0.798 & 0.760\\
Cora-full & 0.767 & 0.753 & 0.717\\
Lastfm & 0.687 & 0.683 & 0.688\\
\bottomrule
\end{tabular}
\end{table}

We further investigate whether the two key assumptions made by our attacks (see \autoref{subsection:ThreatModel}) can be relaxed: 1) the adversary has a shadow dataset that comes from the same distribution as the target dataset, 2) the adversary has a shadow model with the same architecture as the target model.

\mypara{Different Shadow Dataset Distribution}
\autoref{table:attack_performance_different_shadow_dataset} shows the attack results when the shadow model is trained on a dataset from a different distribution.
We observe that in this case, our combined attack can still achieve similar or even better performance compared to the same distribution shadow dataset.
For instance, when the target model is GraphSAGE trained on Cora, the attack accuracy is 0.775 with Citeseer as the shadow dataset, while the accuracy is 0.767 with Cora as the shadow dataset.
This shows that even the adversary does not have the same distribution shadow dataset, they can still launch effective membership inference.

\begin{figure}[!t]
\centering
\begin{subfigure}{0.49\columnwidth}
\includegraphics[width=\columnwidth]{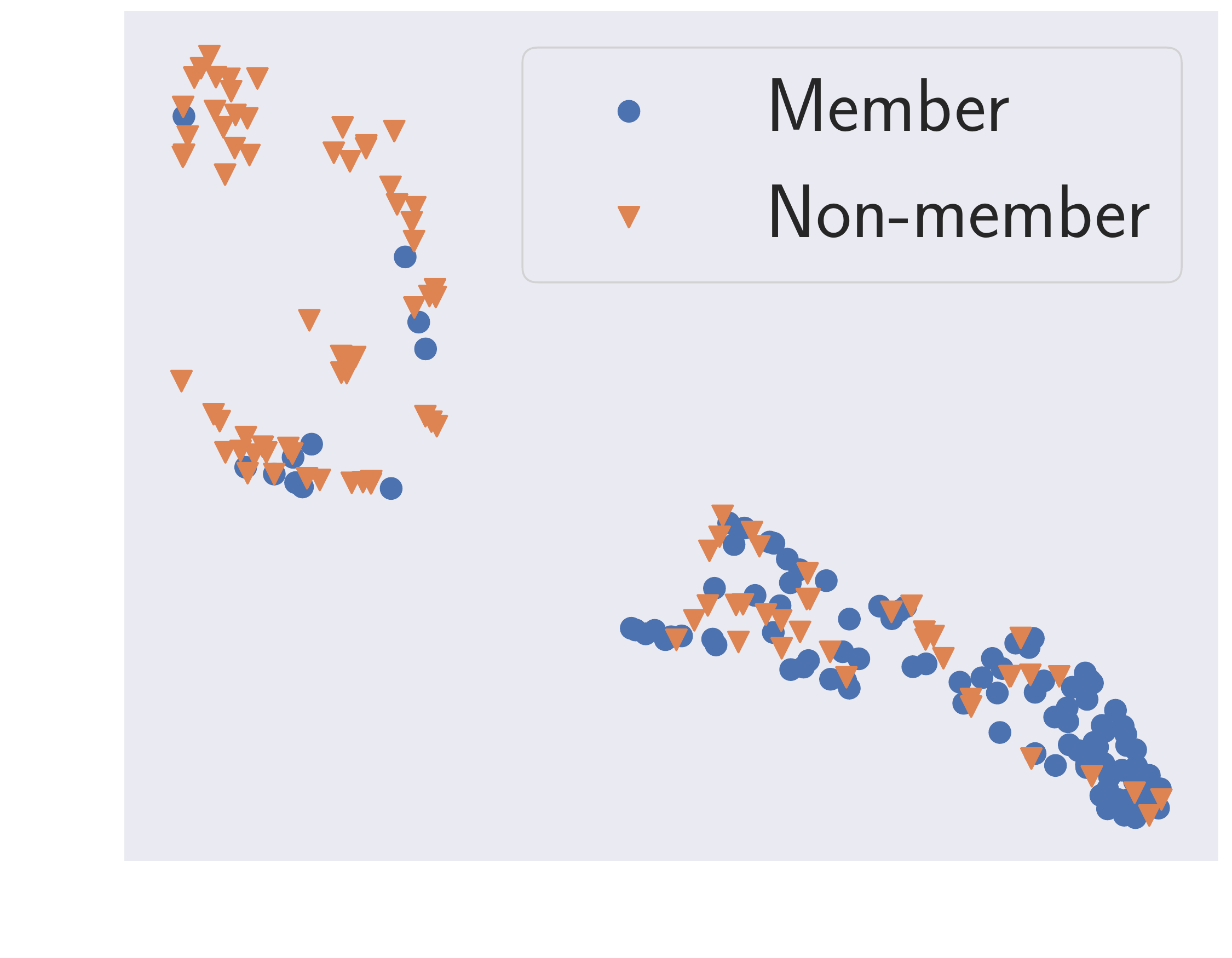}
\caption{Cora as Shadow Dataset}
\label{figure:tsne_Cora_Cora}
\end{subfigure}
\begin{subfigure}{0.49\columnwidth}
\includegraphics[width=\columnwidth]{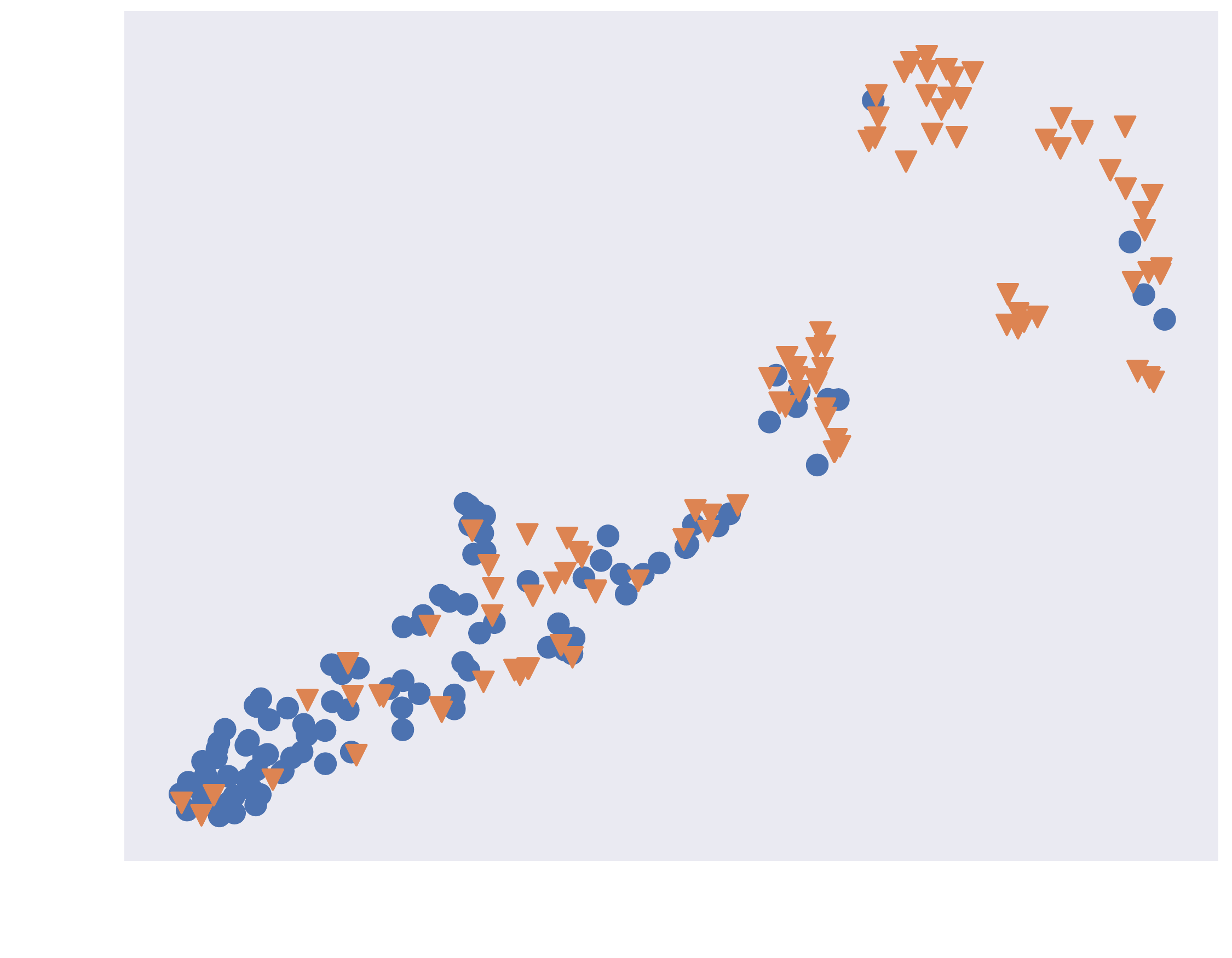}
\caption{Citeseer as Shadow Dataset}
\label{figure:tsne_Cora_Citeseer}
\end{subfigure}
\caption{The embeddings of 100 randomly selected member and non-member nodes obtained from the combined attack's hidden layer.
We project them into a 2-dimensional space using t-SNE.
The target model is GraphSAGE trained on Cora.
The two shadow models are GraphSAGE trained on Cora or Citeseer.
Each point represents a node.}
\label{figure:tsne_attack}
\end{figure}

We further extract the embeddings of members and non-members from two combined attack models (one corresponds to the same distribution shadow dataset, the other corresponds to the different distribution shadow dataset), and project the embeddings into a 2-dimensional space using t-SNE~\cite{MH08}.
The results are shown in \autoref{figure:tsne_attack}.
In both cases, member and non-member nodes are easily separable.

\mypara{Different Shadow Model Architecture}
The results of using different shadow model architectures are summarized in \autoref{table:attack_performance_different_shadow_model}.
We see that a shadow model with a different architecture from the target model still yields good attack performance.
For instance, for GraphSAGE trained on Cora-full, the attack accuracy is 0.753 when the shadow model architecture is GIN while the original attack accuracy is 0.767.

\begin{figure}[!t]
\centering
\includegraphics[width=\columnwidth]{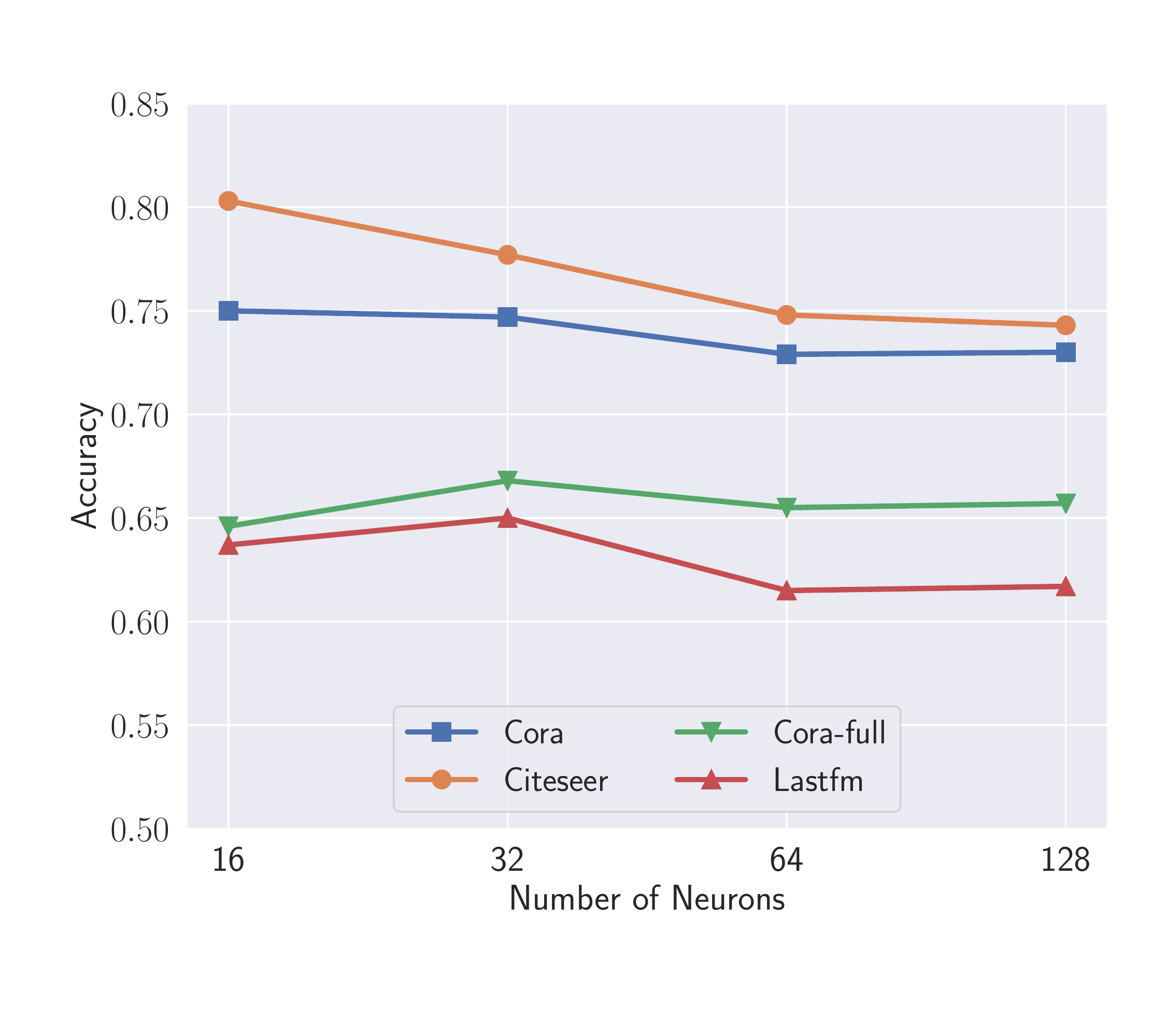}
\caption{The performance of membership inference attacks when the target model is GAT with 32 neurons in its hidden layer and the shadow model is GraphSAGE with different numbers of neurons in its hidden layer.
The x-axis represents the number of neurons in the shadow model's hidden layer.
The y-axis represents membership inference attacks' accuracy.}
\label{figure:different_shadow_neurons}
\end{figure}

We further investigate whether the different number of neurons in the shadow model affects the attack performance.
Concretely, we evaluate the case when the target model is GAT with 32 neurons in its hidden layer, and the shadow model is GraphSAGE with different numbers of neurons ranging from 16 to 128.
The results are depicted in \autoref{figure:different_shadow_neurons}.
We observe that the attack performance is relatively stable under different numbers of neurons.

\medskip
In conclusion, both the same distribution shadow dataset and same architecture shadow model assumptions can be relaxed, which further demonstrates the severe membership privacy risks of GNNs.

% ----------------------------------------------------
\subsection{Possible Defenses}
\label{subsection:Defense}
% ----------------------------------------------------

\begin{figure*}[!t]
\centering
\begin{subfigure}{0.66\columnwidth}
\includegraphics[width=\columnwidth]{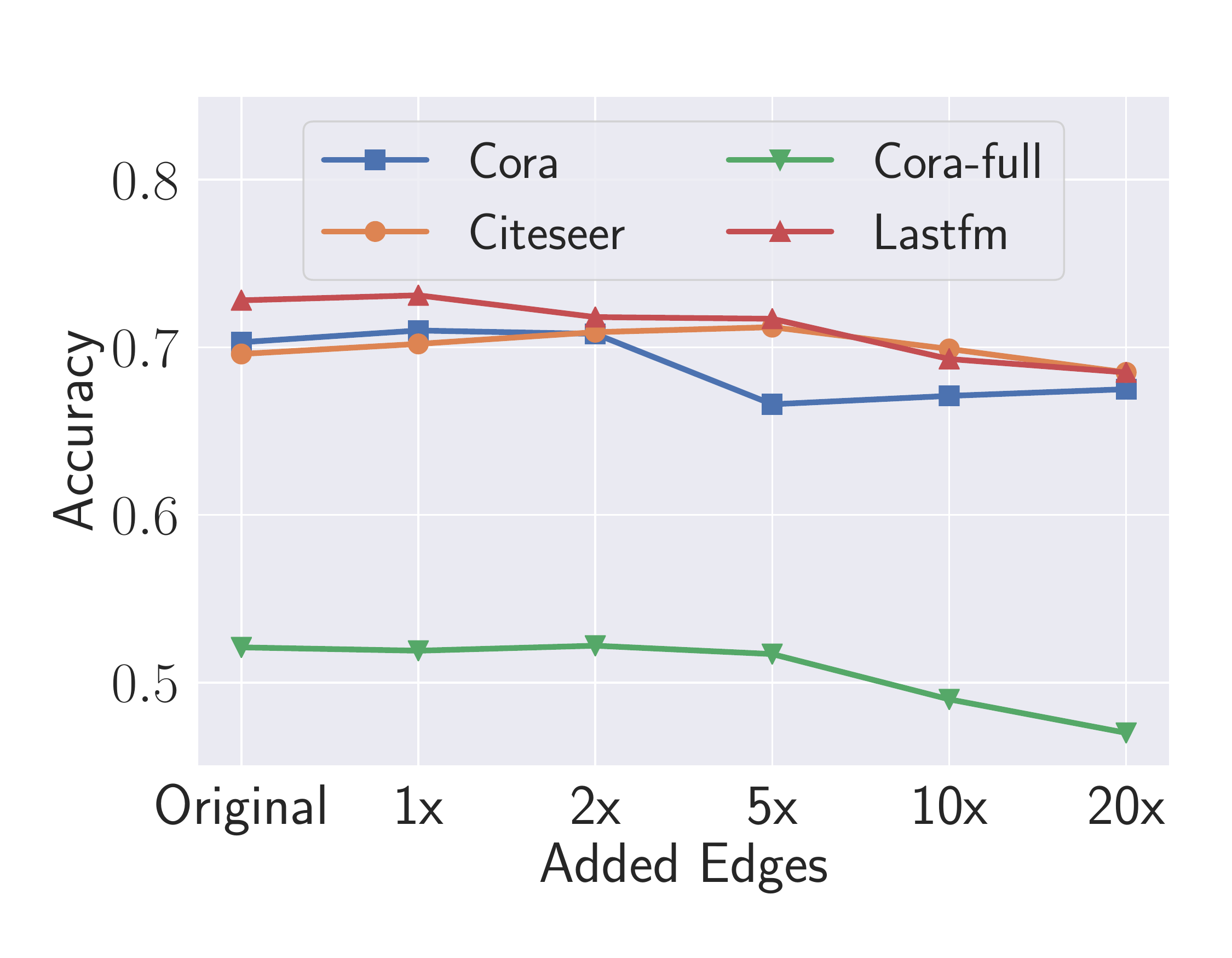}
\caption{Target Model Performance (0-hop)}
\label{figure:defense_add_edge_target_performance_0hop}
\end{subfigure}
\begin{subfigure}{0.66\columnwidth}
\includegraphics[width=\columnwidth]{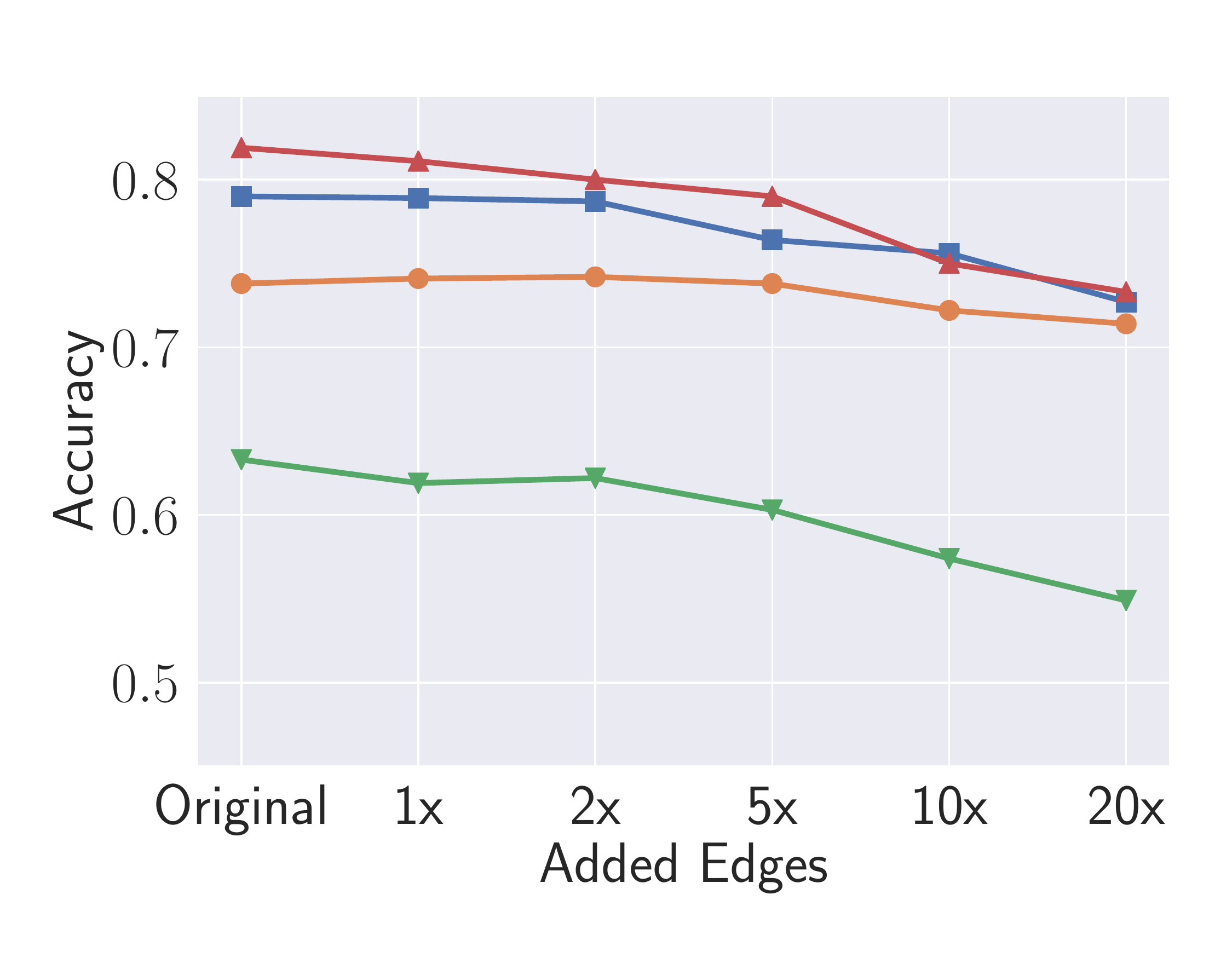}
\caption{Target Model Performance (2-hop)}
\label{figure:defense_add_edge_target_performance_2hop}
\end{subfigure}
\begin{subfigure}{0.66\columnwidth}
\includegraphics[width=\columnwidth]{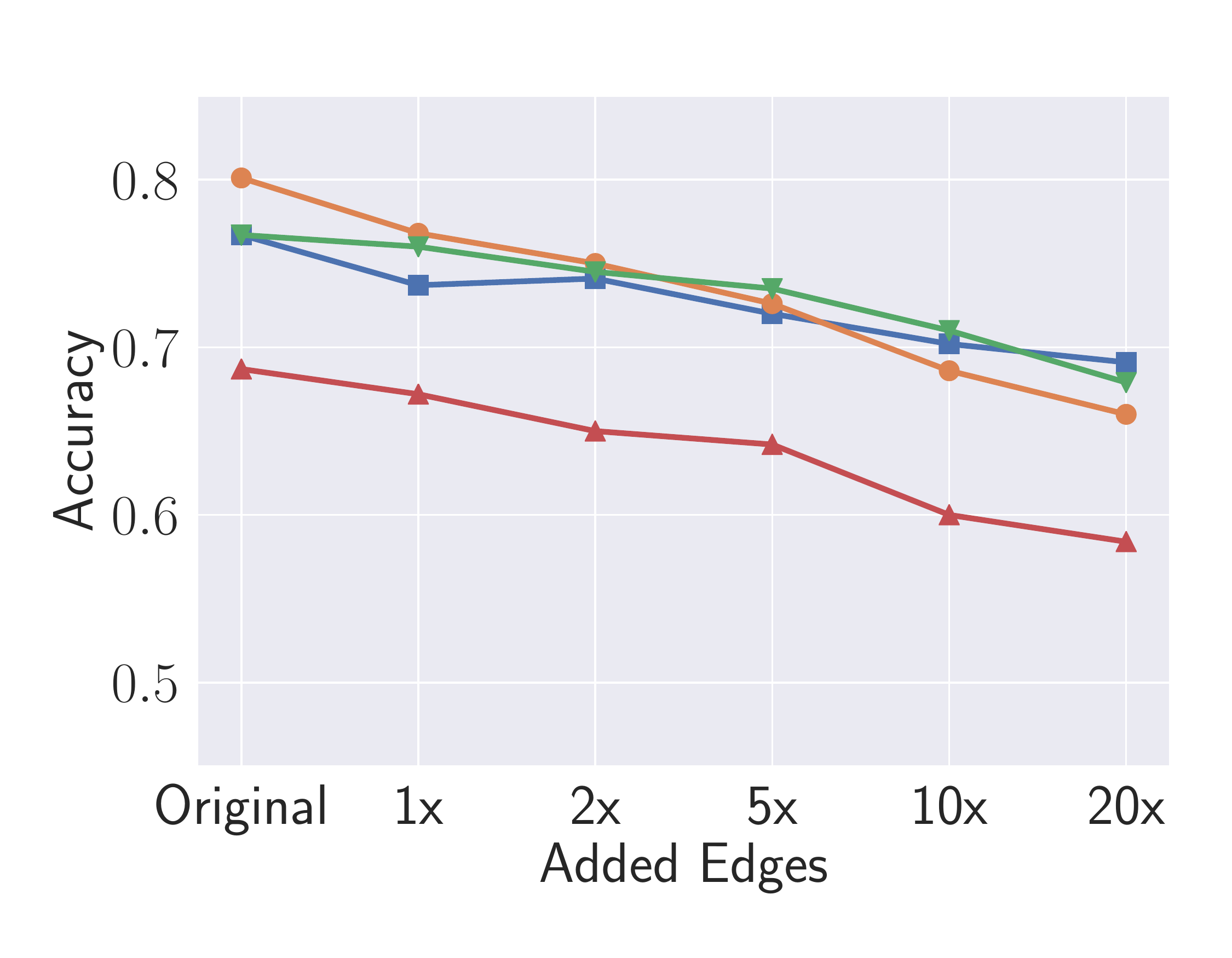}
\caption{Attack Performance ($\AttackModel_{c}$)}
\label{figure:defense_add_edge_attack_performance}
\end{subfigure}
\caption{The performance of the target model's original task and membership inference attacks when applying random edge addition as the defense.
The x-axis represents different proportions of edges added. 
Here, 2$\times$ means randomly adding in total 2 times more edges in the target training dataset.
The y-axis represents the accuracy of the target models' original classification tasks or membership inference attacks.
Note that we only show the results when GraphSAGE is used as the architecture for both target and shadow models.}
\label{figure:defese_add_edge}
\end{figure*}

\begin{figure*}[!t]
\centering
\begin{subfigure}{0.50\columnwidth}
\includegraphics[width=\columnwidth]{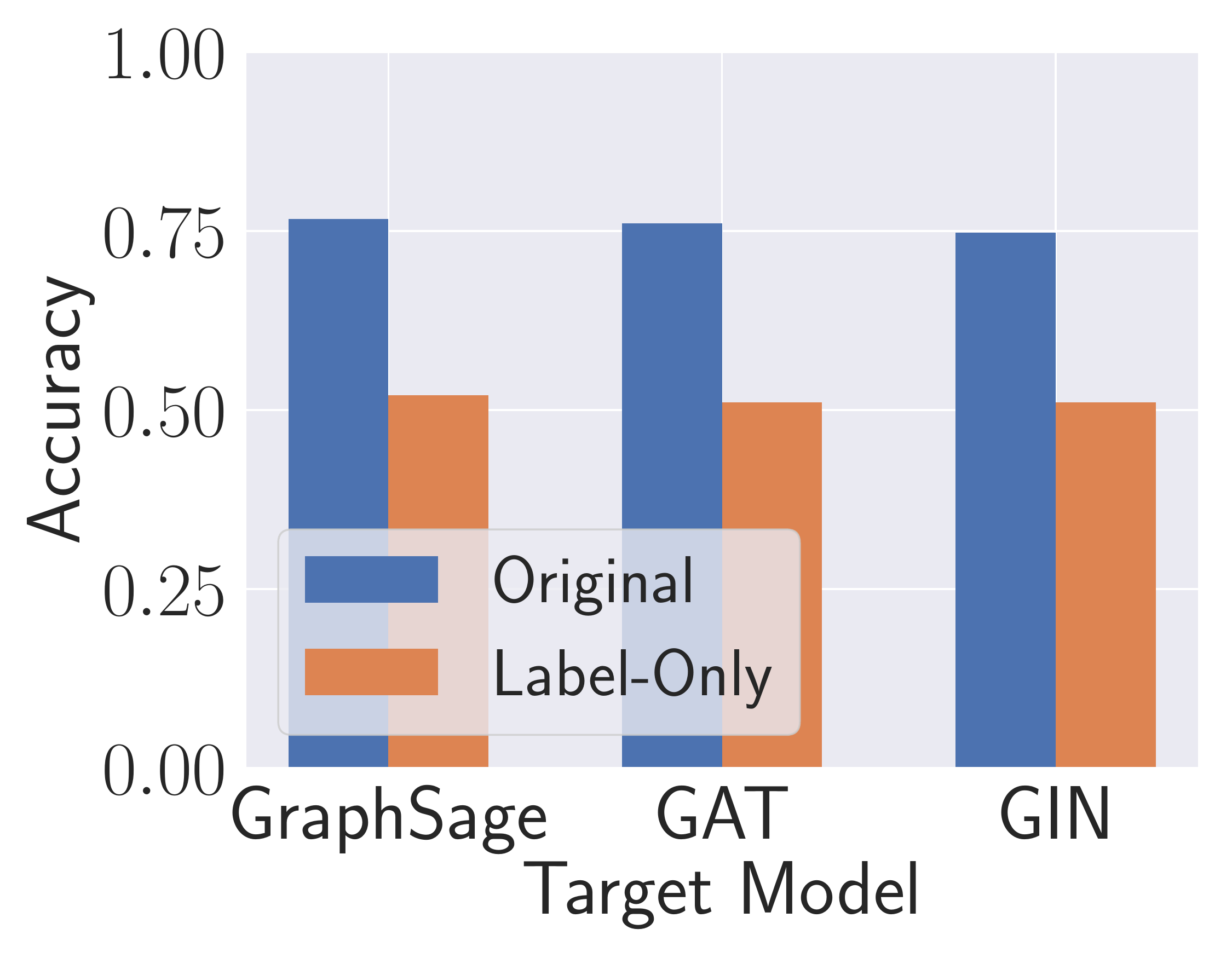}
\caption{Cora}
\label{figure:defense_label_only_attack_performance_Cora}
\end{subfigure}
\begin{subfigure}{0.50\columnwidth}
\includegraphics[width=\columnwidth]{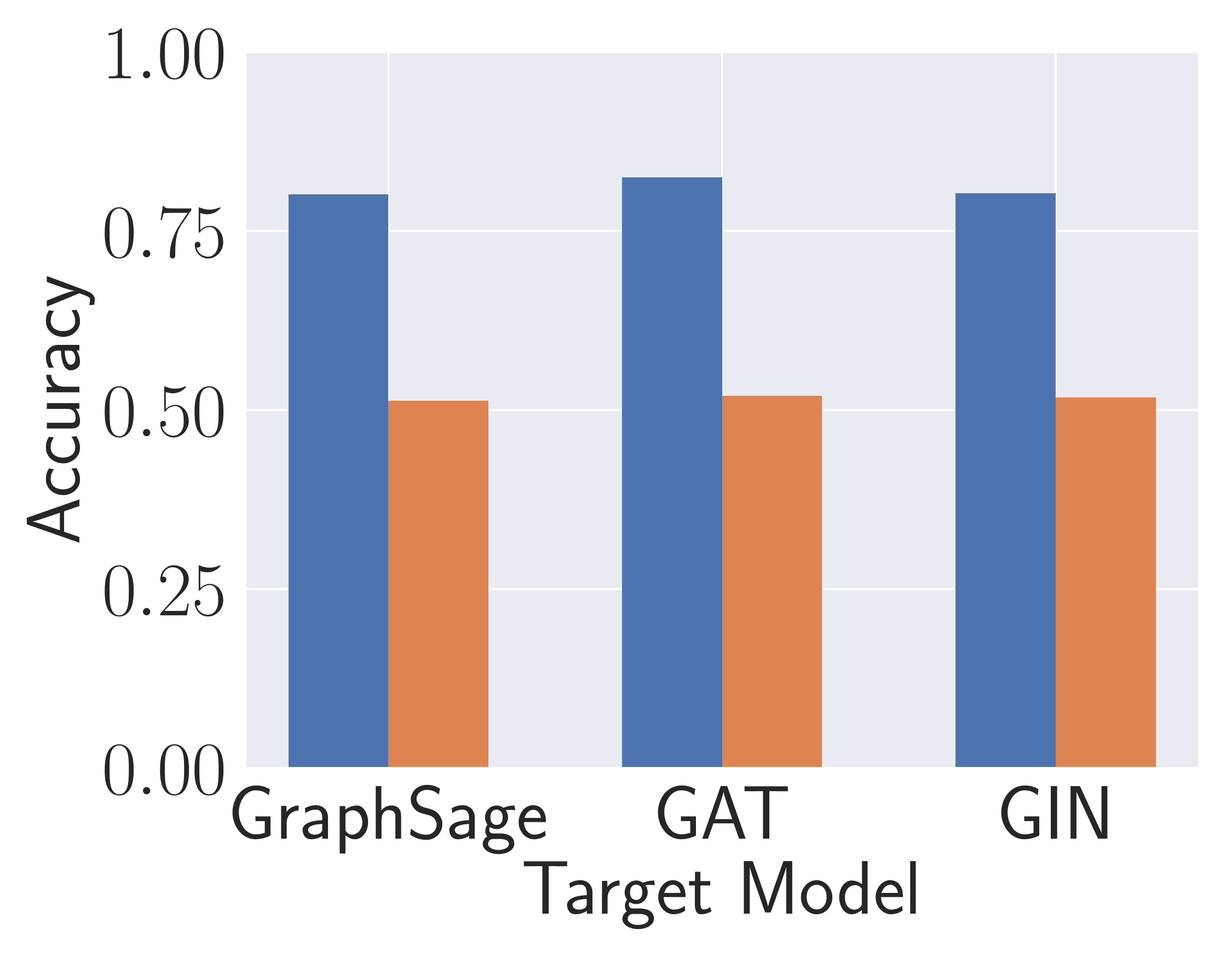}
\caption{Citeseer}
\label{figure:defense_label_only_attack_performance_Citeseer}
\end{subfigure}
\begin{subfigure}{0.50\columnwidth}
\includegraphics[width=\columnwidth]{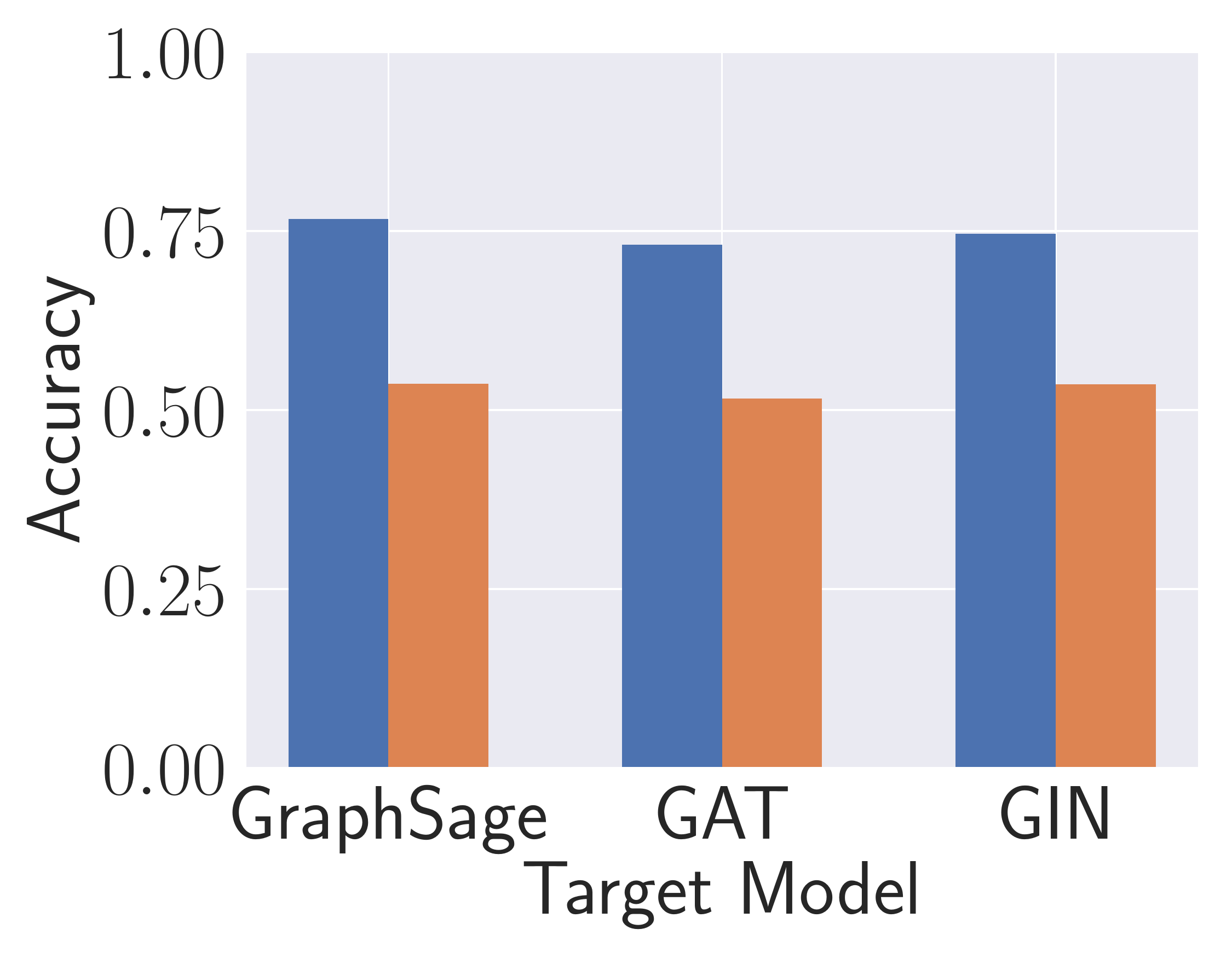}
\caption{Cora-full}
\label{figure:defense_label_only_attack_performance_Corafull}
\end{subfigure}
\begin{subfigure}{0.50\columnwidth}
\includegraphics[width=\columnwidth]{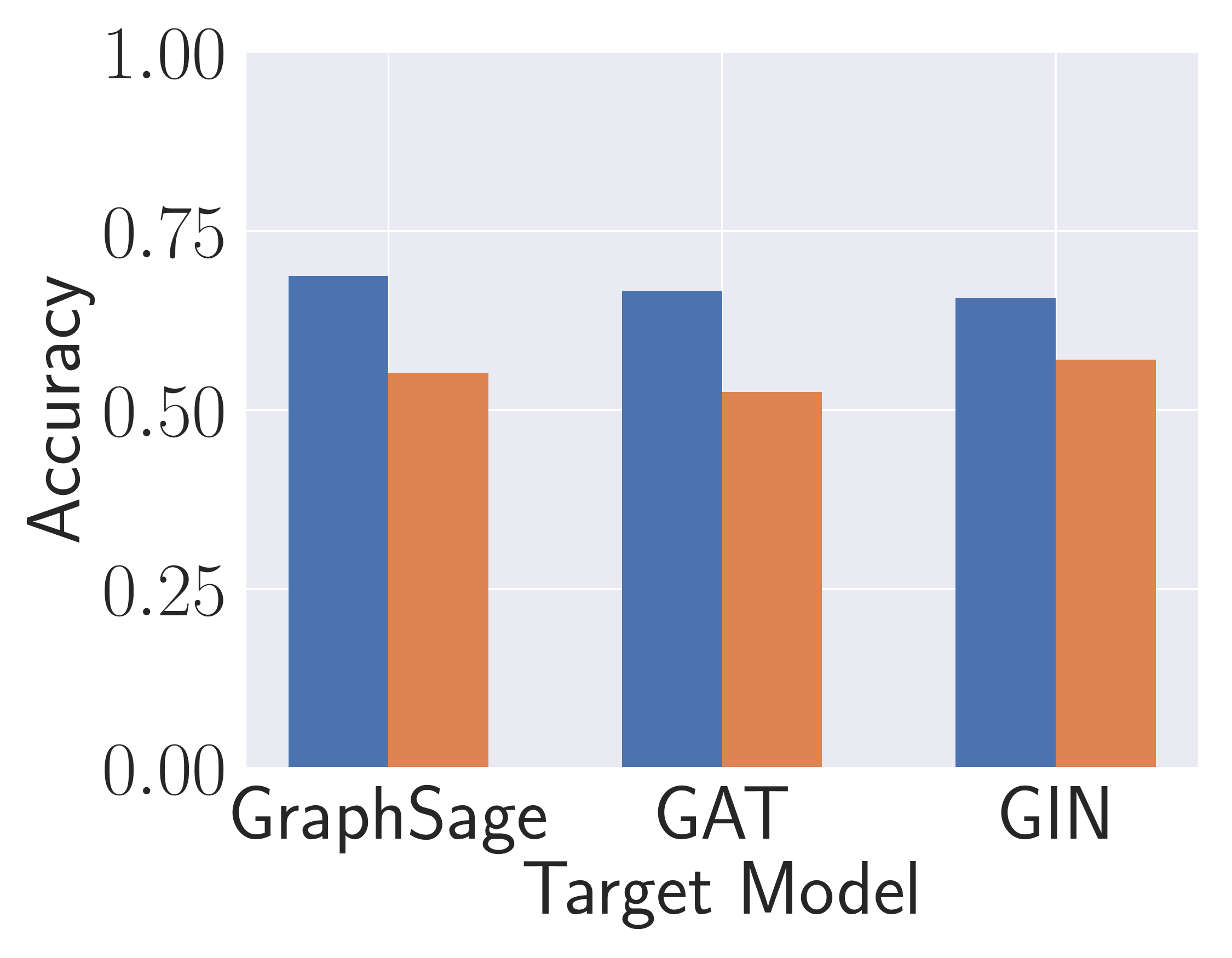}
\caption{Lastfm}
\label{figure:defense_label_only_attack_performance_lastfm}
\end{subfigure}
\caption{The performance of membership inference attacks when applying label-only output as the defense.
The x-axis represents different target models' architectures.
The y-axis represents the accuracy of membership inference attacks.}
\label{figure:defense_label_only_attack_performance}
\end{figure*}

To mitigate the membership inference attacks, we investigate two possible defense mechanisms, namely random edge addition and label-only output.

\mypara{Random Edge Addition} 
In the first defense, we perturb the target training dataset's graph structure by randomly adding edges.
For the adversary, the shadow model is trained on the original shadow training dataset.
We evaluate the target models' performance with respect to the original classification task, i.e., utility, and the membership inference attack performance using the combined attacks.
Due to space limitations, we only show the results when both the target and shadow models use GraphSAGE as their architecture in \autoref{figure:defese_add_edge}.
Other models exhibit similar trends.

In \autoref{figure:defense_add_edge_attack_performance}, we observe that with more random edges added, the attack performance indeed drops.
For instance, the membership inference accuracy is 0.801 on the original Citeseer dataset, while the accuracy drops to 0.660 when 20 times more edges are added.
This indicates that adding random edges to the target training dataset can protect nodes' membership privacy.
As shown in \autoref{figure:micro_degree_vs_auc}, nodes with higher degree suffer less membership leakage risks, since the aggregation function of GNN during training aggregates more neighbor nodes' information and ``memorize'' less about the target node itself.
On the other hand, the target models also suffer large utility loss as shown in \autoref{figure:defense_add_edge_target_performance_0hop} and \autoref{figure:defense_add_edge_target_performance_2hop}.
For instance, the accuracy of original classification task is 0.819 on the original Lastfm dataset using 2-hop query, while the accuracy decreases to 0.733 when 20 times more edges are added (the corresponding attack accuracy drops from 0.687 to 0.584).

\mypara{Label-Only Output} 
For the second defense, we let the target model only return the prediction label instead of posteriors.
In this case, we assume that the adversary knows the total number of classes of the target model.
The adversary first converts the prediction labels derived from the 0-hop and 2-hop queries into two one-hot vectors, respectively.
Then, the two vectors serve as the input to train the combined attack model $\AttackModel_{c}$.

The performance of membership inference attacks against different target models is shown in \autoref{figure:defense_label_only_attack_performance}.
We observe that on all the target models, membership inference attack accuracy decreases significantly.
For instance, on Cora-full, when both the target and shadow models' architectures are GraphSAGE, the membership inference accuracy of the original combined attack is 0.767, while the accuracy drops to 0.537 after applying the label-only output defense.
In addition, this defense can also limit the target model's utility as labels contain less information than posteriors.
We note that previous work~\cite{CTCP20,LZ20} investigate the label-only membership inference attack on non-GNN models.
However, its effectiveness on the proposed defense for GNN models remains unjustified and we leave it as our future work.

\medskip
In summary, our proposed defense mechanisms can reduce the risk of membership inference attack against GNN models.
However, they both limit the target model's utility.
In the future, we plan to investigate more advanced defense mechanisms.

% ----------------------------------------------------
\section{Related Work}
\label{section:RelatedWork}
% ----------------------------------------------------

\mypara{Membership Inference Attack}
Membership inference attacks aim at inferring membership of individual training samples of a target model to which an adversary has black-box access through a prediction API~\cite{SSSS17,SZHBFB19,NSH18,YGFJ18,HMDC19,NSH19,CXXLBKZ18,SS202,CLEKS19,LZ20}.
Most of the existing attacks focus on deep learning models that are trained on sensitive data from the Euclidean space, such as images and texts.
Shokri et al.~\cite{SSSS17} propose the first membership inference attack against machine learning models in the black-box setting.
The authors provide a general formulation of membership inference attack whereas the adversary trains multiple shadow models to mimic the target model's behavior with certain background knowledge of training data and leverages many attack models to conduct the attack.
Salem et al.~\cite{SZHBFB19} further relax several key assumptions from~\cite{SSSS17}, such as knowledge of the target model architecture, shadow dataset from the same distribution.
Yeom et al.~\cite{YGFJ18} discuss the relationship between overfitting and membership attacks.
Nasr et al.~\cite{NSH19} conduct a comprehensive study for membership inference attacks in both black-box and white-box settings.
To mitigate the attacks, some defense mechanisms~\cite{SSSS17,SZHBFB19,NSH18,JSBZG19} have been proposed. 
Those strategies include using model stacking~\cite{SZHBFB19}, dropout~\cite{SZHBFB19}, adversarial training~\cite{NSH18}, jointly
maximize privacy and prediction accuracy~\cite{JSBZG19}, etc.

\mypara{Other Exploratory Attacks Against ML Models}
Besides membership inference, other exploratory attacks such as model inversion, attribute inference, and model stealing have been studied by many researchers.
In model inversion attacks~\cite{FLJLPR14,FJR15,ZJPWLS20}, an adversary aims to reconstruct input samples from a target ML model, i.e., model inversion enables the adversary to directly learn information about the training dataset.
Fredrikson et al.~\cite{FLJLPR14} first propose a model inversion attack in the setting of drug dose classification.
Later, Fredrikson et al.~\cite{FJR15} further extend model inversion to general ML settings relying on back-propagation. 
More recently, Zhang et al.~\cite{ZJPWLS20} develop a more advanced attack based on GANs to synthesize the training dataset.

Attribute inference attacks~\cite{SS20,MSCS19} aim to infer some general properties of the training dataset. 
Meils et al.~\cite{MSCS19} first show that collaborative machine learning can leak sensitive attributes about training data.
Song and Shmatikov~\cite{SS20} later demonstrate that having risks of attribute inference is an intrinsic feature of machine learning models, which is caused by overlearning.

The goal of model stealing attacks~\cite{TZJRR16,WG18,OSF19,JCBKP20} is to extract the parameters from a target model.
Tramer et al.~\cite{TZJRR16} propose the first model stealing attack, with black-box access to the target model.
Wang and Gong~\cite{WG18} propose hyperparameters stealing attacks for a variety of ML models, based on the observation that the gradient of the objective function at the value of parameters is close to 0.
More recently, Orekondy et al.~\cite{OSF19} propose a model stealing attack based on reinforcement learning, which relaxes assumptions on dataset and model architecture.

\mypara{Adversarial Attacks Against Graph Neural Networks}
Recent studies show that GNNs are susceptible to privacy and security attacks~\cite{ZAG18,BG192,DLTHWZS18,ZG19,WWTDLZ19,WG19,ZJWG20}.
Most of these attacks against GNNs are causative attacks where the adversary can manipulate the training dataset by introducing adversarial perturbations to node features, graph structure, etc.
In this direction, different adversarial attack strategies have been investigated. 
Z{\"u}gner et al.~\cite{ZAG18} design adversarial examples on attributed graphs. 
Bojchevski et al.~\cite{BG192} construct poisoning attacks on unsupervised node embeddings based on random walks. 
Wang and Gong~\cite{WG19} perform the study on an adversarial attack for collective classification. 
Z{\"u}gner and G{\"u}nnemann~\cite{ZG19} introduce training time attacks on GNNs. 

The emerging GNN models~\cite{HYL17,XHLJ19, VCCRLB18} enable the adversary to launch exploratory attacks.
Unlike the causative attacks, these attacks do not intend to change the parameters of the target models.
Instead, the adversary probes the target models with carefully crafted input data and learn from the responses.
Though exploratory attacks on classical machine learning models have been extensively studied~\cite{TZJRR16,WG18,OSF19,JCBKP20,SS20,MSCS19,FLJLPR14,FJR15,ZJPWLS20},  only a few studies focus on exploratory attacks on GNNs~\cite{HJBGZ21,WYPY20,DBS20,ONK21}.
He et al.~\cite{HJBGZ21} propose the first link stealing attack to infer whether there is an edge between two nodes used to train the target GNN model.
Wu et al.~\cite{WYPY20} focus on the GCN model extraction attack given various levels of background knowledge.
Duddu et al.~\cite{DBS20} and Olatunji et al.~\cite{ONK21} have performed some preliminary studies on node-level membership inference attacks against GNNs.
However, Duddu et al.~\cite{DBS20} lack a clear attack methodology.
Olatunji et al.~\cite{ONK21} only conduct attacks in a restricted scenario, i.e., using a target node's 2-hop subgraph to query the target model to obtain the input to their attack model, which is deficient to provide a complete picture of the membership inference risks stemming from GNNs.
Also, compared to these work, we further perform an in-depth analysis of the factors that influence the attack performance, such as ego density and feature similarity, and investigate two defenses.
We refer the audience to other works~\cite{SDYWYHL18,DLTHWZS18,CLPXCXHZ20,JLXWT20,XMLDLTJ20} for comprehensive surveys of existing adversarial attacks and defenses on GNNs. 

% ----------------------------------------------------
\section{Conclusion}
\label{section:Conclusion}
% ----------------------------------------------------

In this paper, we perform a comprehensive privacy risk assessment of graph neural networks through the lens of node-level membership inference attacks.
We systematically define the threat model along three dimensions, including shadow dataset, shadow model, and node topology, and propose three different attack models.
We conduct extensive experiments on three popular GNN models over four benchmark datasets.
Our evaluation results show that GNNs are indeed vulnerable to membership inference attacks even with minimal background knowledge of an adversary.
Moreover, our analysis reveals that a node's degree, ego density, and feature similarity have a large impact with respect to the attack performance.
We further show that the attacks are still effective even the adversary does not have the same distribution shadow dataset or same architecture shadow model.
To mitigate the attacks, we propose two possible defense mechanisms and discuss their trade-offs between membership privacy and model utility.

% ----------------------------------------------------
\bibliographystyle{plain}
\bibliography{normal_generated_py3}
% ----------------------------------------------------

% ----------------------------------------------------
\end{document}